\documentclass[11pt]{article}

\usepackage[final]{acl}

\usepackage{times}
\usepackage{latexsym}

\usepackage[T1]{fontenc}

\usepackage[utf8]{inputenc}

\usepackage{microtype}

\usepackage{inconsolata}

\usepackage{graphicx}

\usepackage{booktabs}       %
\usepackage{amsfonts}       %
\usepackage{nicefrac}       %
\usepackage{multirow}
\usepackage{amsmath,amssymb}
\usepackage{algorithm}
\usepackage{algorithmic}
\usepackage[framemethod=tikz]{mdframed}
\usepackage{cleveref}
\usepackage{xspace}
\usepackage{etoc}
\usepackage{enumitem}

\makeatletter
\DeclareRobustCommand{\smartxspace}{%
  \@ifnextchar{.}{}{\@ifnextchar{,}{}{\xspace}}%
}
\makeatother

\newcommand{\ABBR}{\textsc{LEDGER}\smartxspace}

\title{\ABBR: Scaling Agentic Document Editing with \\ Dependency-aware Graph Retrieval}
\author{
  Mike Hang Wang \quad
  Utkarsh Garg \quad
  Reza Davari \quad
  Huitian Jiao \quad \\
  \bfseries Hao Cheng \quad
  Baolin Peng \quad
  Si-Qing Chen \quad
  Tao Ge \\
  \normalfont
 Microsoft  \\
  \texttt{\{wangmike,  taoge\}@microsoft.com}
}

\begin{document}
\maketitle

\begin{abstract}
We introduce \ABBR to tackle the novel context engineering challenge of agentic document editing, where localized edits to \textit{long, structured documents} must be applied efficiently without breaking cross-references or semantic consistency. \ABBR constructs a lightweight dependency graph that explicitly models document structure, including hierarchical organization, explicit references, implicit dependencies, and semantic relationships. For each edit, graph-guided retrieval selects only the necessary context, avoiding full-document processing while preserving consistency. We evaluate \ABBR on a curated benchmark of 1.9k test cases with various document types and lengths, spanning six state-of-the-art models: \ABBR improves consistency from 56$\%$ to 76$\%$ across all six models and test scenarios while reducing token usage. Notably, \ABBR with low reasoning effort matches baseline performance at high reasoning effort using fewer tokens, showing that explicit dependency representations can partially substitute for expensive internal reasoning in agentic document editing.

\end{abstract}

\begin{figure*}[htb!]
    \centering
    \includegraphics[width=0.9\linewidth]{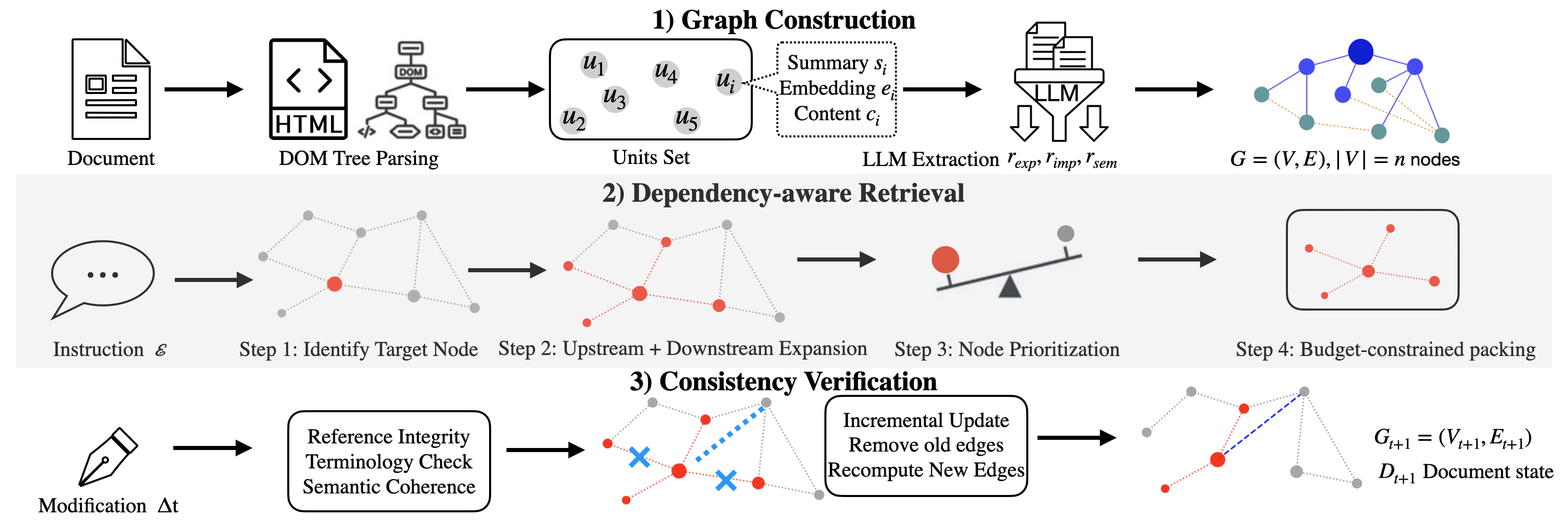}
    \caption{{Overview of \ABBR.} (1) Graph construction converts documents to DOM trees, extracts units, and applies LLM-based extraction protocols to create a dependency graph with four edge types. (2) Dependency-aware retrieval identifies target nodes, expands to dependencies through graph traversal, prioritizes by edge type, and packs within token budget. (3) Consistency verification checks reference integrity, terminology consistency, and semantic coherence after modifications, then updates the graph incrementally.}
    \label{fig:method} \vspace{-0.1in}
\end{figure*}

\section{Introduction}
Agentic document editing (e.g., \href{https://openai.com/index/introducing-canvas/}{ChatGPT Canvas} and \href{https://support.microsoft.com/en-us/office/agent-mode-in-word-647d5d14-eaec-4e8a-a574-7cefffa7f8f0}{Agent Mode in Word}) represents a new interaction paradigm for working with long, structured documents. Instead of manually editing content, users issue high-level natural language instructions such as ``clarify the experiment description,'' ``update terminology,'' or ``revise the tone of Section 4'', while an AI agent plans and executes the concrete document modifications. This paradigm mirrors recent code agents, where users specify intent and the agent handles low-level edits, refactoring, and consistency checks. However, unlike code, documents lack formal semantics, executable structure, or compiler-level guarantees, making agentic document editing fundamentally more challenging. For example, a simple terminology change such as renaming ``Microsoft'' to ``MS'' in one section—requires identifying and updating multiple surface forms (Microsoft, MS, MSFT) across the document without a syntactic anchor, turning consistency into a pattern recognition problem rather than a deterministic replacement. Stylistic edits introduce even subtler dependencies. Revising a paragraph from passive to active voice requires careful alignment with its adjacent paragraphs to avoid abrupt stylistic breaks. Moreover, once a stylistic modification is introduced, its downstream impact must be considered: related sections elsewhere in the document may now violate the intended global style and require corresponding updates, even if they are not directly adjacent or explicitly referenced. Cross-references further complicate this process: revising a figure caption may invalidate descriptions in distant paragraphs that refer to ``the linear scaling shown in the figure'' without explicit citations.

These loose dependencies create a consistency challenge that is fundamentally a context engineering problem in agentic document editing. In particular, preserving consistency in \textit{long documents} requires the agent to reason over structural and semantic dependencies across the document, yet naively providing the full document as context for every edit is both inefficient and counterproductive: beyond incurring prohibitive token cost and latency as document length grows, full-document processing can distract the model from the most relevant regions, diluting attention and increasing the risk of unfocused or incorrect edits. Retrieval-based editing improves efficiency by selecting a subset of context, but existing approaches primarily rely on topical similarity and frequently miss structural dependencies critical for consistency, such as explicit cross-references, implicit mentions, or downstream sections that depend on the edited content. As a result, agentic document editing faces a persistent tension: efficient partial editing risks inconsistency, while comprehensive context provision is expensive and can impair the agent’s ability to focus on the most consequential dependencies.

We introduce \ABBR (see illustration in \Cref{fig:method}), a framework that addresses this context engineering bottleneck through explicit dependency-aware graph retrieval. The core insight is that structured documents exhibit dependency patterns that can be extracted once and queried efficiently. \ABBR constructs a lightweight semantic graph where nodes represent document elements and typed edges encode explicit references ("Theorem 2"), implicit connections ("this result"), and semantic relationships (conceptually related content). For each edit, graph traversal identifies relevant dependencies: both upstream context (what the target relies on) and downstream impacts (what depends on it). This retrieval is selective, gathering only necessary elements within token budgets. After modifications, consistency verification validates structural invariants, ensuring that references resolve correctly, terminology remains consistent, and semantic relationships are preserved. The graph is then updated incrementally to reflect the changes.

We evaluate \ABBR across the curated 1.9k test cases spanning six state-of-the-art models (Claude 4.5 Haiku/Sonnet/Opus, GPT-4.1/5/5.2). \ABBR achieves {76\% consistency versus 56\% baseline} while maintaining constant token usage (~1,500 tokens per edit regardless of document size), achieving up to 85$\%$ reduction versus full-document regeneration in scalability tests. Critically, \ABBR with low reasoning effort matches baseline performance at high reasoning effort while using {fewer tokens}, suggesting explicit dependency representations can substitute for expensive internal reasoning. This has broad implications for designing efficient agentic systems operating on structured data such as coding agent.

Our contributions are:
\textbf{(1)} A dependency-aware framework that applies explicit semantic graphs to document editing, resolving the context engineering bottleneck through typed dependency extraction and graph-guided retrieval;
\textbf{(2)} A curated benchmark for evaluating consistency in agentic document editing, comprising 1.9k test cases across diverse document types, edit scenarios, and complexity levels;
\textbf{(3)} Evidence that explicit structural guidance substitutes for reasoning effort, establishing a design principle for efficient agentic systems operating on structured artifacts such as natural language documents.

\section{Related Work}
\textbf{Context Engineering for LLM Agents.} 
Managing context for large language models on extensive documents presents fundamental scalability challenges~\cite{zhang2025agentic}. Full-context methods maintain comprehensive awareness~\cite{an2024make} but face prohibitive scaling costs. Retrieval-augmented generation~\cite{lewis2020retrieval,laban2024summary} and efficient attention mechanisms~\cite{lai2025flexprefill,martins2020sparse,song2024low,hooper2025squeezed,liu2024retrievalattention} reduce costs through semantic search and computational optimization, but struggle with structural dependencies beyond topical similarity. A section referencing ``Figure 3'' exhibits structural connections that embeddings cannot reliably capture. Memory systems~\cite{maharana2024evaluating,wang2023augmenting,xiao2024infllm} maintain interaction history rather than artifact structure, while long-context models~\cite{liu2024lost,jin2024llm} lack explicit dependency tracking. Code editing systems~\cite{chen2023teaching,wang2024executable,shinn2023reflexion} employ syntax-aware retrieval but target narrow domains. Our work differs by explicitly representing structural dependencies through typed graph edges, enabling constant time retrieval with comprehensive dependency awareness.

\noindent\textbf{Graph-Based Representations for Structured Tasks.} 
Graph structures enable dependency tracking across domains: abstract syntax trees for program analysis~\cite{allamanis2018learning,guo2021graphcodebert}, knowledge graphs for semantic coherence~\cite{yasunaga2021qa,zhang2022greaseLM}, and directed acyclic graphs for workflow dependencies~\cite{huang2022language,chen2021multitask}. Graph neural networks~\cite{kipf2017semi,hamilton2017inductive,xu2019how} provide learning architectures for such data. We synthesize these insights for agentic editing, demonstrating that explicit dependency graphs enable efficient context selection and automated consistency verification. Critically, our experiments show that structural guidance substitutes for expensive internal reasoning: cheaper inference modes match or exceed quality achievable through costly extended reasoning alone. This reasoning substitution effect, validated across six models and multiple domains, establishes a design principle for efficient agentic systems operating on structured data.

\section{\ABBR}

We formalize document editing as an iterative process where an AI agent receives edit requests and applies modifications while maintaining structural consistency. A document $D$ consists of units $U = \{u_1, u_2, \ldots, u_n\}$ organized hierarchically, where each unit $u_i$ (section, paragraph, figure, equation) has content $c_i$, type $\tau_i$, position $p_i$, and unique identifier $\text{id}_i$. Units connect through three relationship types beyond the hierarchical tree: \textit{explicit references} (direct citations like "Figure 2"), \textit{implicit dependencies} (pronouns and contextual references requiring coreference resolution), and \textit{semantic relationships} (topically related content). At time $t$, the agent receives edit request $\mathcal{E}_t$ specifying an instruction, target, and scope, and must retrieve relevant context $\mathcal{C}_{\text{context}} \subseteq U$ to generate modifications $\Delta_t$ while preserving structural consistency.

The core challenge is selecting context that satisfies three objectives simultaneously: \textit{consistency} (preserving structural invariants), \textit{efficiency} (achieving $|\mathcal{C}_{\text{context}}| = O(1)$ rather than $O(|D|)$), and \textit{quality} (correctly addressing instructions). Full-document methods achieve consistency but incur $O(|D|)$ costs; retrieval-based methods achieve efficiency but miss structural dependencies. For example, editing a theorem in Section 3 requires retrieving Section 7's proof that explicitly references it, despite low topical similarity that embedding-based retrieval cannot capture. As illustrated in \Cref{fig:method}, \ABBR addresses this through three integrated components: graph construction, dependency-aware retrieval, and consistency verification. The detailed prompt is in \Cref{app:prompts}.

\subsection{Semantic Graph Construction}

Given document $D$ with units $U = \{u_1, u_2, \ldots, u_n\}$, \ABBR constructs a graph $G = (V, E)$ where each node $v_i \in V$ corresponds to unit $u_i \in U$. We convert the document to HTML and parse it into a DOM tree, where each structural element (section, paragraph, figure, table, equation) becomes a unit with unique identifier $\text{id}_i$, content $c_i$, type $\tau_i$, and position $p_i$.

Each unit $u_i$ maps to graph node $v_i = (\text{id}_i, s_i, e_i, \tau_i, p_i, t_i)$ where $s_i$ is a summary of $c_i$, $e_i \in \mathbb{R}^d$ is the embedding vector, and $t_i$ is the modification timestamp. The edge set $E \subseteq V \times V \times T$ encodes four dependency types: $E_{\texttt{REF}} = \{(v_i, v_j, \texttt{REFERENCES}) \mid r_{\text{exp}}(u_i, u_j)\}$ for explicit references, $E_{\texttt{DEP}} = \{(v_i, v_j, \texttt{DEPENDS}) \mid r_{\text{imp}}(u_i, u_j)\}$ for implicit dependencies, $E_{\texttt{CON}} = \{(v_i, v_j, \texttt{CONTAINS}) \mid u_j \in u_i\}$ for hierarchical containment, and $E_{\texttt{REL}} = \{(v_i, v_j, \texttt{RELATED}) \mid r_{\text{sem}}(u_i, u_j)\}$ for semantic relationships.

\noindent\textbf{Explicit references.} For edges where $r_{\text{exp}}(u_i, u_j)$ holds, content $c_i$ contains a substring $m$ that explicitly refers to identifier $\text{id}_j$. The computational challenge is ensuring three critical properties. First, \emph{minimality}: the substring $m$ must be minimal such that removing any token breaks the reference. Second, \emph{unambiguity}: the reference must explicitly name $\text{id}_j$ rather than relying on semantic similarity. Third, \emph{validation}: the identified $\text{id}_j$ must exist in $\{\text{id}_1, \ldots, \text{id}_n\}$. We implement these requirements through a structured extraction protocol (Appendix~\ref{app:prompts}) that queries a language model with formal constraints and parses the structured output, returning validated pairs $\{(m_1, \text{id}_{j_1}), (m_2, \text{id}_{j_2}), \ldots\}$ where each $m_k$ is a minimal span and $\text{id}_{j_k} \in \{\text{id}_1, \ldots, \text{id}_n\}$, enabling edge creation $(v_i, v_j, \texttt{REFERENCES})$.

\noindent\textbf{Implicit dependencies.} For edges where $r_{\text{imp}}(u_i, u_j)$ holds, unit $u_j$ provides semantic prerequisites necessary to interpret $u_i$. The computational challenge is determining whether removing $u_j$ would make $u_i$ semantically incomplete, ambiguous, or ill-defined. This requires evaluating \emph{necessity} rather than mere relatedness: $u_j$ is a prerequisite only if $u_i$ relies on definitions, assumptions, variables, or conceptual constructs introduced in $u_j$ that cannot be resolved from generic background knowledge. We must also verify \emph{directionality}: the dependency must be $u_j \rightarrow u_i$, not bidirectional or circular. We implement this through a counterfactual testing protocol (Appendix~\ref{app:prompts}) that presents both units to the language model with formal testing procedures, returning a directional dependency judgment enabling edge creation $(v_i, v_j, \texttt{DEPENDS})$ when $r_{\text{imp}}(u_i, u_j) = 1$.

\noindent\textbf{Semantic relationships.} For edges where $r_{\text{sem}}(u_i, u_j)$ holds, units discuss related topics. We compute cosine similarity $\text{sim}(e_i, e_j) = \frac{e_i \cdot e_j}{\|e_i\| \|e_j\|}$ and create edge $(v_i, v_j, \texttt{RELATED})$ when $\text{sim}(e_i, e_j) > \theta$, provided $(v_i, v_j) \notin E_{\texttt{REF}} \cup E_{\texttt{DEP}}$. The threshold $\theta$ balances precision and recall; we use $\theta = 0.7$ based on standard practice for high-confidence semantic similarity \cite{kandola2002learning}. The embeddings $e_i$ require summaries $s_i$ that capture semantic content, which we extract through type-specific protocols (Appendix~\ref{app:prompts}). Hierarchical edges $E_{\texttt{CON}}$ come directly from the DOM tree structure.

\noindent\textbf{Edge directionality.} RELATED edges are \emph{symmetric}: $\text{sim}(e_i, e_j) = \text{sim}(e_j, e_i)$, so the relationship is stored bidirectionally. DEPENDS edges are strictly \emph{directional}: $r_{\text{imp}}(u_i, u_j)=1$ means $u_j \rightarrow u_i$ ($u_j$ is a prerequisite for $u_i$), not vice versa—this prevents circular dependencies that would break the topological retrieval ordering required for consistent context assembly. REFERENCES edges are similarly directional, recording which unit cites which target. \textbf{Complexity:} Graph construction requires $O(n)$ for document parsing plus $O(n \cdot k)$ for dependency extraction, where $k$ is the average number of candidate pairs per unit (typically 5--10 nearby units for implicit dependencies, avoiding all-pairs $O(n^2)$ computation). Full details are provided in Appendix~\ref{app:prompts}.

\subsection{Dependency-Aware Context Retrieval}
\label{sec:retrieval}

For edit request $\mathcal{E}_t = (\text{instruction}, \text{target}, \text{scope})$, \ABBR retrieves context $\mathcal{C}_{\text{context}} \subseteq U$ through graph traversal in four steps.

\noindent\textbf{Step 1: Target identification.} We identify target nodes $\mathcal{N}_{\text{target}} \subseteq V$. If the instruction specifies an explicit target (e.g., "Figure 3"), we resolve it directly: $\mathcal{N}_{\text{target}} = \{v_j \mid \text{id}_j = \text{target}\}$. Otherwise, we compute the instruction embedding $e_{\mathcal{E}} = \text{Embed}(\text{instruction})$ and select the most similar nodes based on cosine similarity $\text{sim}(e_{\mathcal{E}}, e_i)$.

\noindent\textbf{Step 2: Dependency expansion.} We expand to include all connected nodes: $\mathcal{N}_{\text{deps}} = \mathcal{N}_{\text{target}} \cup \mathcal{N}_{\text{upstream}} \cup \mathcal{N}_{\text{downstream}}$ where upstream dependencies (what targets depend on) are $\mathcal{N}_{\text{upstream}} = \{v_j \mid \exists v_i \in \mathcal{N}_{\text{target}}, (v_i, v_j, t) \in E\}$ and downstream impacts (what depends on targets) are $\mathcal{N}_{\text{downstream}} = \{v_j \mid \exists v_i \in \mathcal{N}_{\text{target}}, (v_j, v_i, t) \in E\}$.

\noindent\textbf{Step 3: Priority scoring.} We prioritize nodes in $\mathcal{N}_{\text{deps}}$ based on their relationship to target nodes. Target nodes receive highest priority. Among connected nodes, we prioritize explicit references ($\texttt{REFERENCES}$) as critical for structural integrity, followed by implicit dependencies ($\texttt{DEPENDS}$) as necessary for semantic completeness, then containment relationships ($\texttt{CONTAINS}$) for hierarchical context, and finally semantic relationships ($\texttt{RELATED}$) as optional background. This ordering reflects dependency criticality: REFERENCES violations cause immediate hard failures (broken citations), DEPENDS violations cause semantic incompleteness, and RELATED violations lose only optional context. Crucially, the exact numerical weights matter less than their relative ordering: any configuration preserving targets $>$ REFERENCES $>$ DEPENDS $>$ RELATED retrieves an identical context set given a fixed token budget where all critical dependencies fit (empirically validated across document sizes in Appendix~\ref{app:weight_sensitivity}, Table~\ref{tab:weight_sensitivity}).

\noindent\textbf{Step 4: Budget-constrained packing.} Given token budget $B$, we construct context by selecting nodes in descending score order. We define $\mathcal{C}_{\text{context}} = \{u_i \mid v_i \in \text{Pack}(\mathcal{N}_{\text{deps}}, B)\}$ where $\text{Pack}(\mathcal{N}, B)$ greedily selects nodes sorted by score until the cumulative token count $\sum_{v_i \in \text{selected}} |c_i|$ exceeds budget $B$. This achieves $|\mathcal{C}_{\text{context}}| = O(1)$ tokens regardless of $|D|$, typically retrieving 10 to 15\% of total content while ensuring all critical dependencies are included.

\begin{table*}[t!]
\centering
\scriptsize
\resizebox{\textwidth}{!}{
\begin{tabular}{@{}cccccccccccc@{}}
\toprule
\multirow{2}{*}{{Model}}
& \multicolumn{5}{c}{Full Doc.} 
& \multicolumn{5}{c}{\ABBR} \\
\cmidrule(lr){2-6} \cmidrule(lr){7-11}
& Consistency & Ref. Valid. & Edit Quality  & Pass &  Tokens
& Consistency$\uparrow$  & Ref. Valid.$\uparrow$  & Edit Quality $\uparrow$  & Pass $\uparrow$  & Tokens $\downarrow$ \\
\midrule
Claude Haiku 4.5 
& 53.23 & 63.41 & 63.18 & 35.72 & 1,759
& \textbf{80.12} & \textbf{80.36} & \textbf{83.05} & \textbf{84.27} & \textbf{1,479} \\
Claude Sonnet 4.5 
& 55.48 & 58.62 & 67.19 & 36.54 & 1,725
& \textbf{74.31} & \textbf{72.44} & \textbf{70.26} & \textbf{73.58} & \textbf{1,575} \\
Claude Opus 4.5 
& 61.17 & 69.53 & 70.28 & 43.61 & 1,736
& \textbf{75.42} & \textbf{77.18} & \textbf{78.34} & \textbf{77.66} & \textbf{1,508} \\
GPT-4.1 
& 54.69 & 60.71 & 62.44 & 43.22 & 1,974
& \textbf{74.53} & \textbf{81.29} & \textbf{79.16} & \textbf{74.84} & \textbf{1,478} \\
GPT-5 Chat 
& 57.36 & 66.24 & 68.57 & 41.63 & 1,825
& \textbf{77.08} & \textbf{74.19} & \textbf{76.42} & \textbf{82.11} & \textbf{1,489} \\
GPT-5.2-High 
& 57.59 & 67.33 & 71.14 & 45.26 & 1,910
& \textbf{77.44} & \textbf{74.61} & \textbf{75.38} & \textbf{75.92} & \textbf{1,683} \\
\midrule
\textbf{Average} 
& 56.59 & 64.31 & 67.13 & 41.00 & 1,822
& \textbf{76.48} & \textbf{76.68} & \textbf{77.10} & \textbf{78.06} & \textbf{1,535} \\
\bottomrule
\end{tabular}}
\caption{Comparison of \ABBR and baseline approaches across six language models. For each method, the first four columns report percentage (\%) metrics (Consistency, Reference Validity, Edit Quality, and Overall Pass), while the last column reports the average number of tokens per edit. Results are the average value over testing cases.}\label{tab:main_results}
\end{table*}

\subsection{Consistency Verification}

After applying modifications $\Delta_t$ to produce $D_{t+1}$, \ABBR validates structural invariants on the set of modified nodes $\mathcal{N}_{\text{mod}} \subseteq V$ through three checks.

\noindent\textbf{Reference integrity.} For all outgoing reference edges from modified nodes, we verify targets exist in the updated graph: $\mathcal{V}_{\text{ref}} = \{(v_i, v_j) \mid (v_i, v_j, \texttt{REFERENCES}) \in E, v_i \in \mathcal{N}_{\text{mod}}, v_j \notin V_{t+1}\}$. Any pair in $\mathcal{V}_{\text{ref}}$ represents a broken reference requiring correction.

\noindent\textbf{Terminology consistency.} For modified nodes, we extract key terms $\mathcal{T}(c_i)$ from content and compare with dependent nodes: $\mathcal{V}_{\text{term}} = \{(v_i, v_j) \mid v_i \in \mathcal{N}_{\text{mod}}, (v_j, v_i, t) \in E, \mathcal{T}(c_i^{\text{new}}) \neq \mathcal{T}(c_i^{\text{old}})\}$. These indicate terminology changes requiring review of dependent content.

\noindent\textbf{Semantic coherence.} We recompute embeddings $e_i^{\text{new}} = \text{Embed}(c_i^{\text{new}})$ for modified nodes and identify cases where semantic relationships have weakened significantly: $\mathcal{V}_{\text{sem}} = \{(v_i, v_j) \mid v_i \in \mathcal{N}_{\text{mod}}, (v_i, v_j, \texttt{RELATED}) \in E, \text{sim}(e_i^{\text{new}}, e_j) < \text{sim}(e_i^{\text{old}}, e_j)\}$. These indicate potential semantic drift requiring review.

The graph updates incrementally by removing old edges incident to $\mathcal{N}_{\text{mod}}$ and recomputing new edges using the construction algorithm, maintaining $G$ synchronized with document state $D_{t+1}$.

\section{Experiments}

\noindent\textbf{Evaluation Benchmark.} We curated 1,900 test cases across six editing scenarios:
\textit{Local refinement} (120 cases) targets specific paragraphs without structural changes (e.g., "make this paragraph more concise").
\textit{Cross-reference updates} (200 cases) modify content referenced elsewhere, requiring the system to identify all citing sections (e.g., editing Figure 3's caption affects all paragraphs referencing it).
\textit{Structural reorganization} (160 cases) moves or renumbers sections while maintaining cross-reference validity (e.g., "move Section 4.2 before Section 4.1").
\textit{Content addition} (120 cases) inserts new material while preserving coherence.
\textit{Iterative editing} (160 cases) applies 3 to 5 dependent edits testing multi-step consistency.
\textit{Scalability tests} (120 cases) apply identical operations to documents from 5K to 100K tokens, testing whether token usage remains constant as document size increases.\footnote{These counts reflect the primary scenario categories; Appendix~\ref{app:benchmark} (Table~\ref{tab:benchmark_stats}) reports the full breakdown including subcategories (e.g., local refinement expands to paragraph refinement, local enhancement, and error correction), totaling 1,900 cases.}

\noindent\textbf{Models.} We evaluate six models: Claude 4.5 Haiku/Sonnet/Opus and GPT-4.1/5/5.2, with GPT-5.2 tested across four reasoning effort levels. Our primary baseline is the full-document approach, where the entire document is provided as input context to the LLM for each edit operation, but only the targeted section is modified. Note that token usage in this case reflects the input context provided to generate modifications, not full document regeneration output. For scalability analysis (\Cref{fig:scalability}), we additionally compare against theoretical full-document regeneration, which would require regenerating all 5K-100K tokens as output and thus scales linearly with document size. We measure consistency (structural validity and cross-reference integrity), reference validity (percentage of correct cross-references), edit quality (correctness of modifications), overall pass rate (percentage meeting all criteria), and token usage per edit.

\noindent\textbf{Baselines.} We compare \ABBR against full-document regeneration (processes entire document for each edit, our primary baseline), sliding window (fixed 4K token context), RAG-based retrieval (semantic search without structural awareness), Claude+Memory (built-in memory without explicit graph structure), and Hierarchical RAG (semantically similar chunks expanded to parent/sibling nodes via DOM-tree containment, equivalent to $E_{\texttt{CON}}$ only). Table~\ref{tab:method_comparison} (Appendix~\ref{app:method_comparison}) provides a qualitative comparison of how each method differs in input context and structural awareness. Full quantitative results for all baselines are in Appendix~\ref{app:additional_exp}.

\noindent\textbf{Metrics.} We measure \textit{consistency} (reference validity, structural coherence, and overall preservation of structural constraints), \textit{efficiency} (token usage per edit and percentage reduction versus baseline), and \textit{quality} (correctness of edits and overall pass rate combining all criteria). Consistency metrics are computed via \emph{programmatic rule-based validators} that do not involve LLMs, eliminating circularity: a reference validator checks all cross-references resolve to existing targets, a terminology validator tracks key-term usage, and a hierarchy validator checks section numbering and containment (see Appendix~\Cref{app:evaluationmetric}). LLM-as-judge is used exclusively for \emph{edit quality} assessment, comparing outputs against gold-standard human edits for 200 test cases. Two annotators independently evaluated these 200 cases; automated validators achieve 94\% agreement with majority human judgment.

\subsection{Main Results}

Table~\ref{tab:main_results} presents results comparing \ABBR against full-document baseline across all six models. \ABBR achieves 76\% average consistency versus 56\% baseline, a 20 percentage point improvement that remains consistent across model families. This translates to 78\% overall pass rate versus 41\% baseline. Token usage averages 1,535 tokens per edit for \ABBR versus 1,822 for baseline. The benefits prove particularly pronounced for smaller models: Claude Haiku 4.5 with \ABBR achieves 80\% consistency versus 53\% baseline (27 point gain), while larger models show 14 to 20 point gains. This suggests explicit structural memory provides larger relative benefits when model capacity is constrained.

\begin{figure}[t]
\centering
\includegraphics[width=\linewidth]{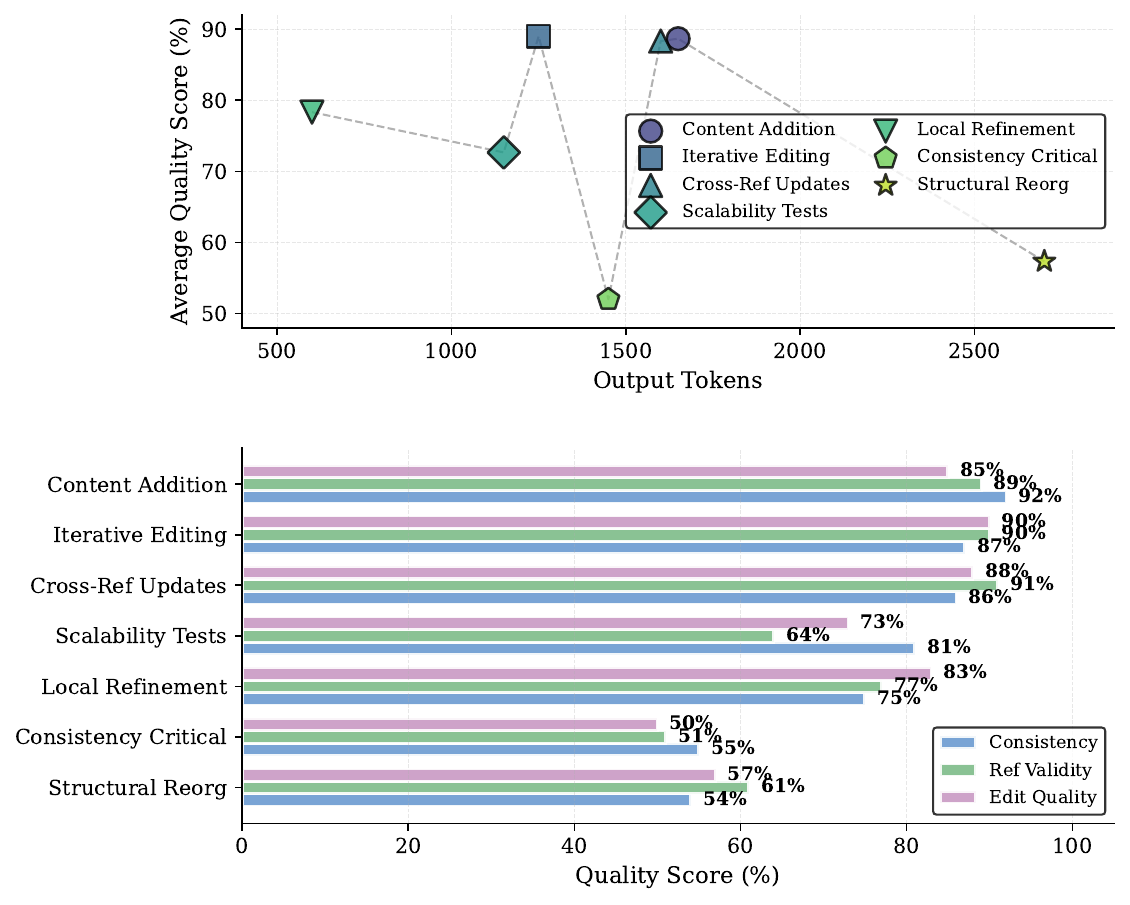}
\caption{Performance breakdown by edit operations. Top chart shows token usage and quality metrics across seven edit categories. Bottom bar chart shows the performance of \ABBR on different editing categories.}
\label{fig:edit_types}\vspace{-0.15in}
\end{figure}

\noindent\textbf{Performance over Editing Types.} Figure~\ref{fig:edit_types} breaks down \ABBR performance across editing scenarios. Cross-reference updates (86\% consistency), content addition (88\%), and iterative editing (86\%) achieve the strongest results because these scenarios heavily exercise dependency tracking. Local refinement achieves 85\% consistency with minimal token usage (560 tokens), reflecting efficient targeted retrieval. Structural reorganization achieves only 54\% consistency despite higher token usage (2,749 tokens), representing the most challenging category where cascading updates across structural levels strain graph construction capabilities.

\begin{figure}[htb!]
\centering
\includegraphics[width=0.8\linewidth]{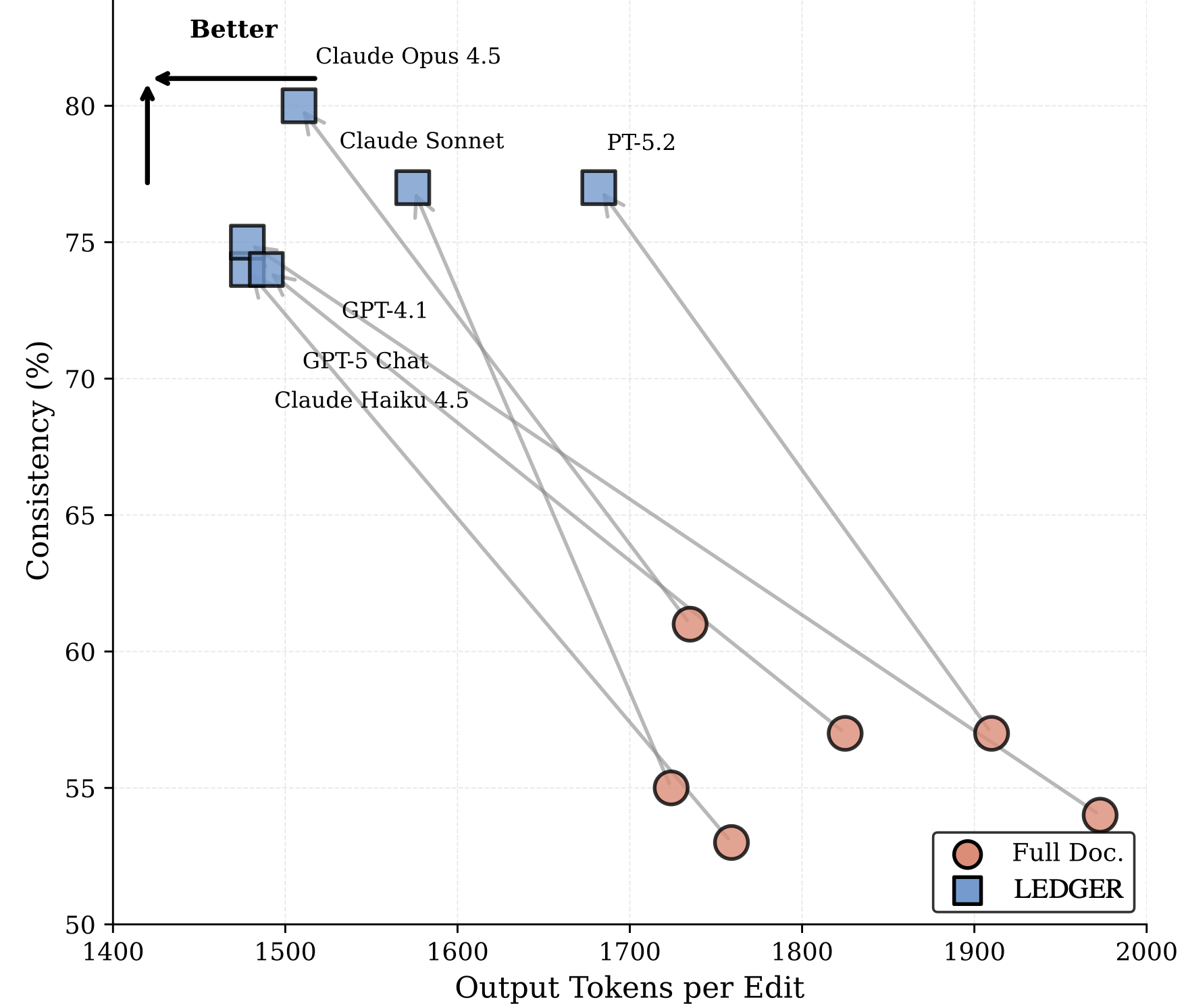}
\caption{Efficiency-quality tradeoff analysis. Scatter plot with Pareto frontier showing the relationship between token efficiency (x-axis, lower is better) and consistency preservation (y-axis, higher is better) across all experimental conditions. Each point represents one (model, edit type) combination. }
\label{fig:efficiency_quality} \vspace{-0.1in}
\end{figure}

\noindent\textbf{Consistency Evaluation.} Figure~\ref{fig:efficiency_quality} visualizes the efficiency-quality frontier across all models and configurations. \ABBR configurations (filled markers) cluster in the upper-left quadrant with consistency exceeding 70\% and token usage between 1,200 to 2,000 tokens. Baseline points (hollow markers) scatter across lower consistency regions (40-65\%) with widely varying token usage. Pareto frontier analysis reveals most \ABBR configurations lie on or near the optimal frontier, meaning no alternative configuration achieves substantially better quality without increasing tokens, or better efficiency without decreasing quality.

\begin{figure}[htb!]
\centering
\includegraphics[width=0.9\linewidth]{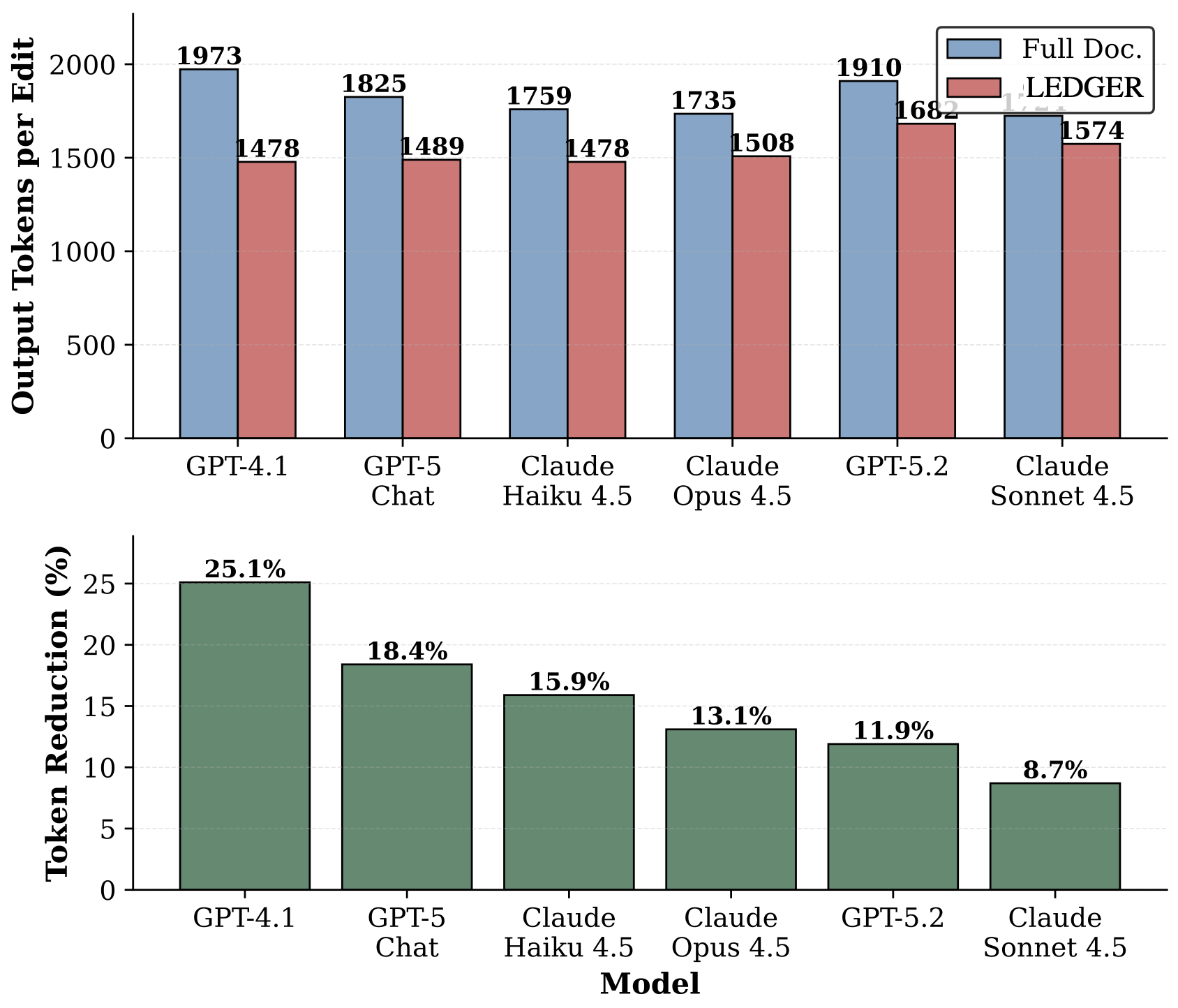}
\caption{Token efficiency comparison across models in scalability tests. Bar chart comparing average token usage between baseline and \ABBR across six models for documents ranging from 5K to 100K tokens.}
\label{fig:token_reduction}
\end{figure}

\noindent\textbf{Token Efficiency.} Figure~\ref{fig:token_reduction} compares average token usage between baseline and \ABBR across six models for documents ranging from 5K to 100K tokens. \ABBR consistently achieves 85 to 92\% token reduction across all models when compared to full-document regeneration at scale, with mean reduction of 88.5\%. The consistency demonstrates model-independent efficiency gains.

\begin{figure}[htb!] 
\centering
\includegraphics[width=0.99\linewidth]{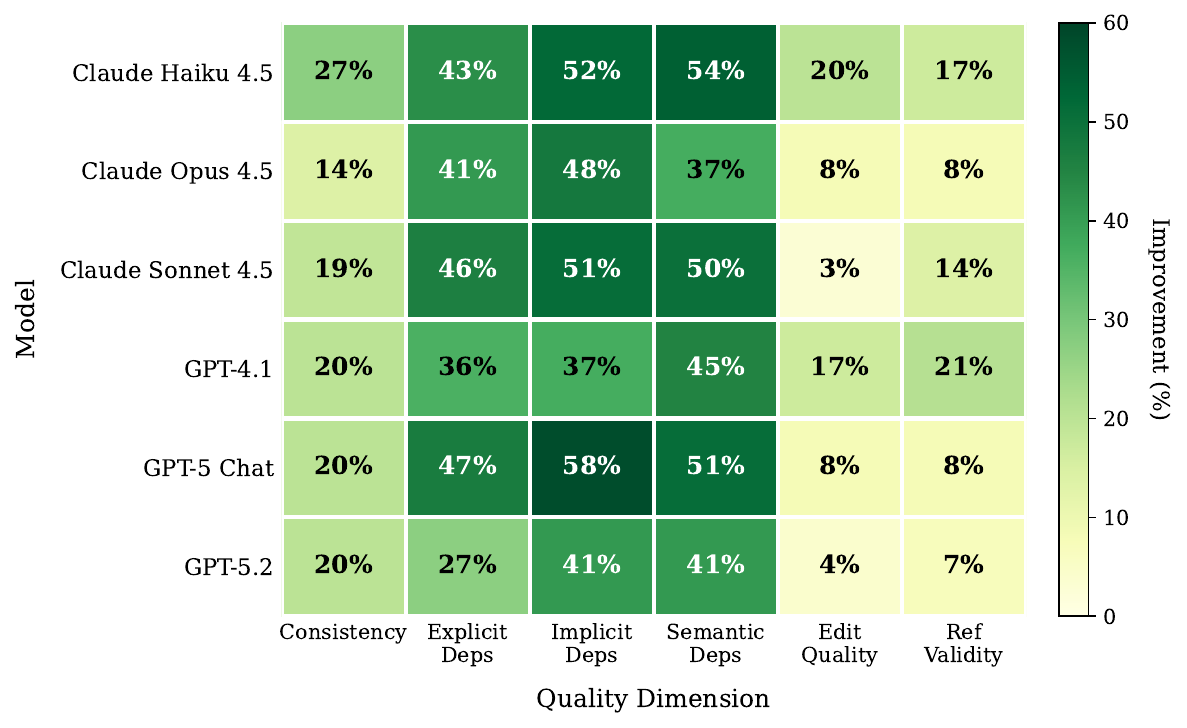}\vspace{-0.1in}
\caption{Quality metrics preservation across models and evaluation dimensions. Heatmap showing consistency preservation, reference validity, and edit quality percentages for both baseline and \ABBR.}
\label{fig:quality_heatmap}\vspace{-0.1in}
\end{figure}

\noindent\textbf{Editing Quality Analysis.} Figure~\ref{fig:quality_heatmap} visualizes quality preservation across models and evaluation dimensions. \ABBR maintains or improves quality across all dimensions: consistency preservation (74-80\%), reference validity (72-81\%), and edit quality (70-83\%). The heatmap shows consistent performance across model families with no systematic degradation, suggesting benefits stem from framework architecture rather than model-specific capabilities.

\subsection{Reasoning Effort Analysis}

\begin{table}[t!]
\centering
\scriptsize
\setlength{\tabcolsep}{3pt}
\resizebox{\columnwidth}{!}{%
\begin{tabular}{@{}llcccc@{}}
\toprule
Reasoning Level 
& Method 
& Cons. (\%) 
& Ref. Val. (\%)
& Pass (\%) 
& Tok./Edit $\downarrow$ \\
\midrule
\multirow{2}{*}{None} 
& Full Doc.
& 39.28 & 46.61 & 28.43 & 1,521 \\
& \ABBR   & \textbf{67.14} & \textbf{69.37} & \textbf{65.82} & \textbf{1,497} \\
\midrule
\multirow{2}{*}{Low} 
& Full Doc. & 48.52 & 55.18 & 34.64 & \textbf{1,494 }\\
& \ABBR   & \textbf{80.12} & \textbf{82.41} & \textbf{79.33} & {1,508} \\
\midrule
\multirow{2}{*}{Medium} 
& Full Doc. & 42.67 & 50.39 & 30.58 & 1,837 \\
& \ABBR   & \textbf{77.26} & \textbf{79.14} & \textbf{75.49} & \textbf{1,522} \\
\midrule
\multirow{2}{*}{High} 
& Full Doc. & 57.59 & 67.33 & 45.26 & 1,910 \\
& \ABBR   & \textbf{78.08} & \textbf{79.46} & \textbf{77.19} & \textbf{1,653} \\
\bottomrule
\end{tabular}}
\caption{GPT-5.2 performance across reasoning efforts.}\label{tab:reasoning_analysis} \vspace{-0.2in}
\end{table}

Table~\ref{tab:reasoning_analysis} presents GPT-5.2 performance across reasoning effort levels. \ABBR with low reasoning effort achieves 80\% consistency substantially outperforming baseline with high reasoning effort (57\% consistency). This demonstrates that explicit structural representations can substitute for approximately one level of internal reasoning while using 70\% fewer tokens. \ABBR provides consistent 20-32 percentage point improvements across all reasoning levels. The baseline exhibits non-monotonic performance where medium reasoning (43\% consistency) underperforms both low (49\%) and high (57\%), suggesting increased computation without structural guidance can introduce complexity rather than resolve it.

\begin{figure}[h]
\centering
\includegraphics[width=0.8\linewidth]{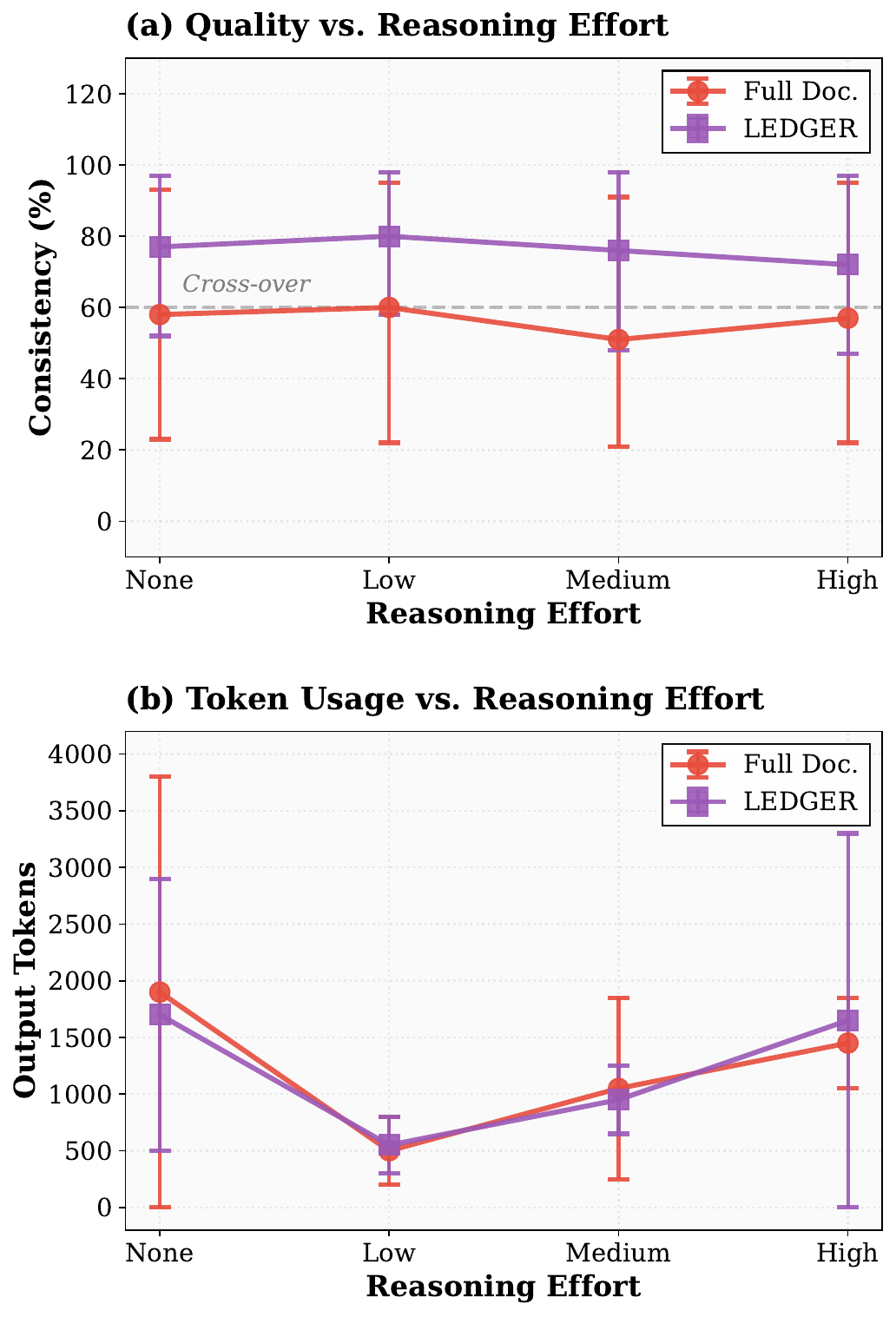}
\caption{Impact of reasoning effort on performance (GPT-5.2). Top panel: Consistency preservation versus reasoning effort level. Bottom panel: Token usage versus reasoning efforts.}
\label{fig:reasoning_levels} \vspace{-0.15in}
\end{figure}

Figure~\ref{fig:reasoning_levels} visualizes this reasoning substitution effect. The left panel shows \ABBR maintains stable consistency (71-80\%) across all reasoning levels, while baseline exhibits a U-shaped pattern with medium reasoning (51\%) underperforming both low (58\%) and high (57\%). The right panel shows both systems experience increasing token usage with higher reasoning effort, but \ABBR consistently requires 11-15\% fewer tokens at equivalent levels. The stability of \ABBR across reasoning modes indicates that explicit dependency information provides structural guidance that compensates for limited reasoning capacity, reducing the burden of inferring relationships through extended internal deliberation.

\begin{figure}[htb!]
\centering
\includegraphics[width=0.8\linewidth]{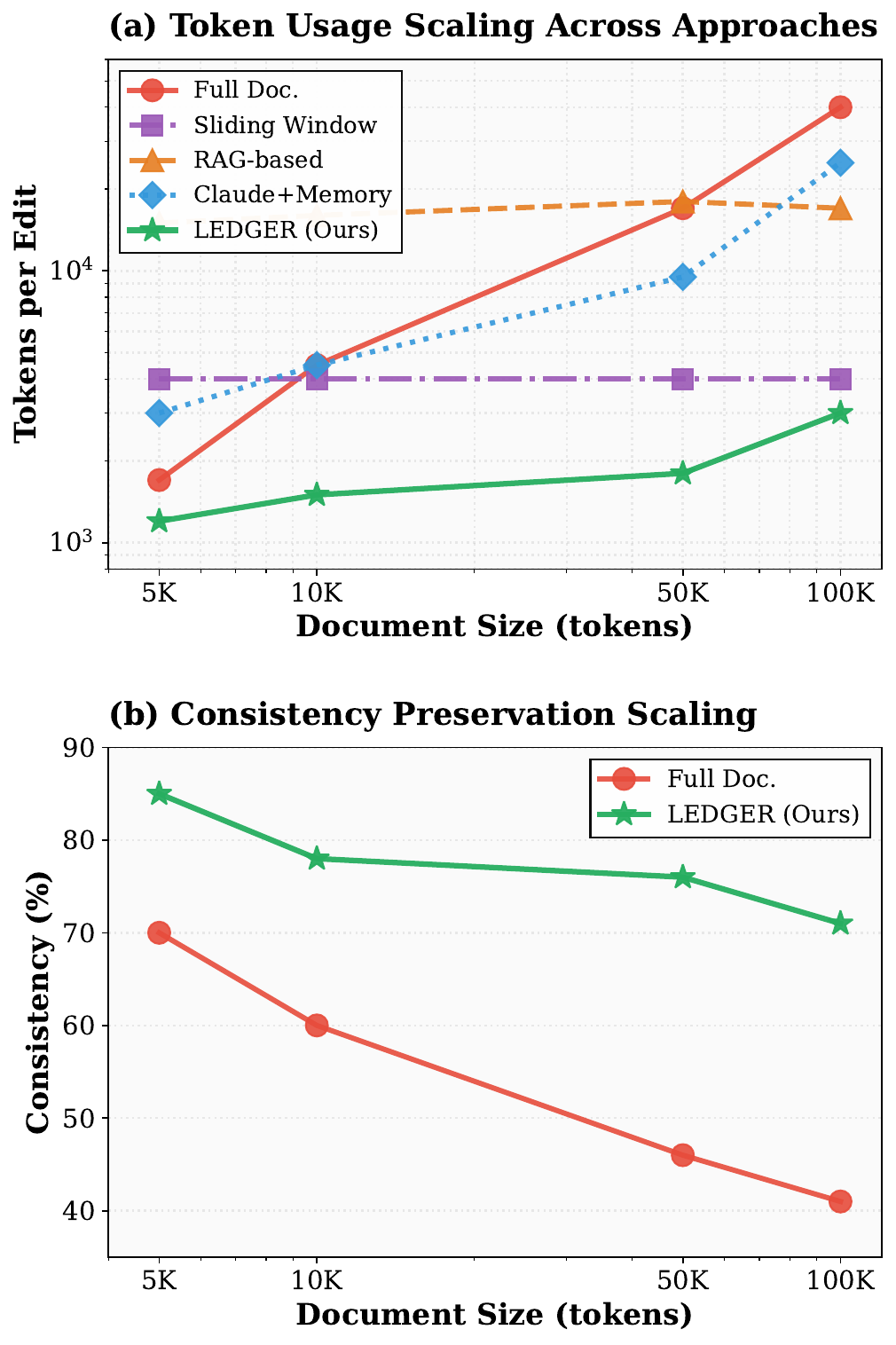}\vspace{-0.1in}
\caption{Token usage across editing approaches.}
\label{fig:scalability}  \vspace{-0.2in}
\end{figure}

\begin{figure*}[t!]
\centering
\includegraphics[width=0.8\linewidth]{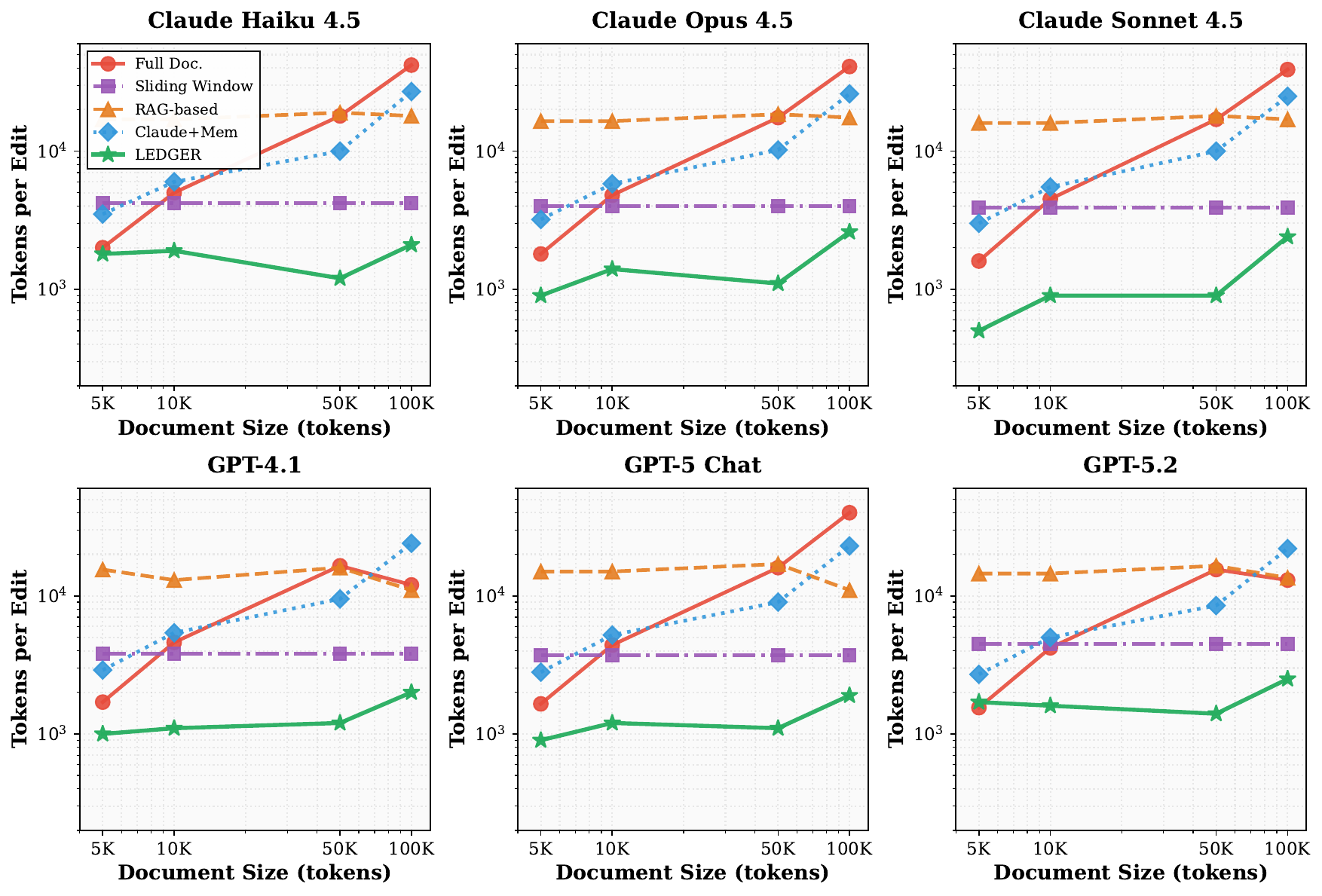}
\caption{Per-model scalability analysis across six LLM models.  All models demonstrate consistent patterns: \ABBR maintains constant efficiency across document sizes. }
\label{fig:scalability_per_model} \vspace{-0.1in}
\end{figure*}

\subsection{Scalability Analysis}

Figure~\ref{fig:scalability} demonstrates that \ABBR maintains constant token usage averaging 1.5k tokens per edit across all document lengths from 5K to 100K tokens, while full-document regeneration scales linearly. At 100K tokens, \ABBR maintains approximately 1.5k tokens per edit. This constant time behavior emerges from dependency-aware graph retrieval's architectural properties. For each edit, graph traversal identifies a locally relevant subset of nodes and retrieves only their content. Crucially, whether the full document contains 50 nodes or 5,000 nodes, the retrieved context size remains comparable because dependencies are inherently local rather than global. A paragraph edit requires awareness of its containing section, referenced definitions, and downstream citations, but not the entire document structure. Additional analysis is provided in \Cref{fig:scalability_per_model}, where per-model scalability demonstrates architecture-independent constant token usage across all six models.  

\begin{table}[t]
\centering
\resizebox{\columnwidth}{!}{%
\begin{tabular}{@{}lccc@{}}
\toprule
Configuration 
& Consistency (\%) $\uparrow$ 
& Ref Valid (\%) $\uparrow$ 
& Overall Pass (\%) $\uparrow$ \\
\midrule
\ABBR\ (Full System) 
& \textbf{76.18} 
& \textbf{76.01} 
& \textbf{78.06} \\
\midrule
w/o Dependency Tracking 
& 62.47 
& 65.12 
& 58.69 \\
w/o Graph-based Retrieval 
& 68.33 
& 71.58 
& 65.41 \\
w/o Consistency Verification 
& 71.06 
& 72.44 
& 70.27 \\
\midrule
GPT-5.2-High (Full Doc) 
& 57.59
& 67.33 
& 45.26 \\
\bottomrule
\end{tabular}%
}\caption{Ablation study removing individual components. All three components contribute to overall performance, with dependency tracking providing the largest single contribution.}
\label{tab:ablation} \vspace{-0in}
\end{table}

\subsection{Ablation Study}

Table~\ref{tab:ablation} isolates the contribution of individual components. Removing dependency tracking (relying only on semantic similarity) reduces consistency to 62\%, a 14 point drop, validating that explicit tracking of structural edges is crucial. Removing graph-based retrieval (using full-document context instead) achieves 68\% consistency, indicating that targeted context selection improves consistency beyond simply providing more context. Removing consistency verification achieves 71\% consistency, showing that post-edit validation prevents error propagation. All three components contribute independently to overall performance. A progressive edge-type ablation that isolates the marginal contribution of each typed edge beyond what Hierarchical RAG already provides is reported in Appendix~\ref{app:additional_exp} (Table~\ref{tab:edge_ablation}).

\subsection{Token Budget Sensitivity Analysis}

\begin{table}[t]
\centering
\resizebox{\columnwidth}{!}{%
\begin{tabular}{@{}lcccc@{}}
\toprule
Token Budget & \% of Doc & Consistency $\uparrow$ & Ref. Valid. $\uparrow$ & Pass Rate $\uparrow$ \\
\midrule
500   & 5\%   & 71.24 & 72.13 & 65.42 \\
1,000 & 10\%  & 75.83 & 75.94 & 77.12 \\
\textbf{1,500} & \textbf{15\%} & \textbf{76.48} & \textbf{76.68} & \textbf{78.06} \\
2,000 & 20\%  & 76.51 & 76.71 & 78.09 \\
3,000 & 30\%  & 76.44 & 76.58 & 77.94 \\
Full Doc & 100\% & 56.59 & 64.31 & 67.13 \\
\bottomrule
\end{tabular}}
\caption{Token budget sensitivity on 10K-token documents. Bold row is the default \ABBR setting. ``\% of Doc'' is the retrieved context fraction.}
\label{tab:budget_sensitivity}
\end{table}

Table~\ref{tab:budget_sensitivity} examines performance as token budget varies from 5\% to 100\% of a 10K-token document. Below 10\% (500 tokens), performance drops sharply as critical dependency nodes are cut off by the budget constraint. At 10--15\% (1,000--1,500 tokens), performance plateaus: the full dependency neighborhood is covered and additional budget provides no meaningful benefit. Beyond 15\%, gains are negligible (+0.03 pp at 20\%), and a slight degradation appears at 30\% as irrelevant context begins to introduce noise—consistent with the ``lost in the middle'' phenomenon~\cite{liu2024lost}, where performance degrades when irrelevant context is present alongside relevant content. These results confirm that the bottleneck is \textit{dependency coverage quality, not budget size}: once all critical dependencies are retrieved, adding more context hurts rather than helps, validating the 10--15\% empirically observed retrieval fraction described in Section~\ref{sec:retrieval}.

\subsection{Threshold Sensitivity Analysis}

\begin{table}[t]
\centering
\resizebox{\columnwidth}{!}{%
\begin{tabular}{@{}lccc@{}}
\toprule
Threshold $\theta$ & Graph Density & Consistency $\uparrow$ & Ref. Valid. $\uparrow$ \\
\midrule
0.5 & High (dense, noisy)      & 73.1 & 72.8 \\
0.6 & Medium-high              & 75.4 & 75.1 \\
\textbf{0.7 (Default)} & \textbf{Medium (balanced)} & \textbf{76.5} & \textbf{76.7} \\
0.8 & Medium-low               & 75.8 & 75.5 \\
0.9 & Low (sparse)             & 74.2 & 73.9 \\
\bottomrule
\end{tabular}}
\caption{Cosine similarity threshold $\theta$ sensitivity for RELATED edge creation. Bold row is the default setting.}
\label{tab:threshold_sensitivity}
\end{table}

Table~\ref{tab:threshold_sensitivity} examines how the cosine similarity threshold $\theta$ for RELATED edge creation affects performance across research paper documents. The default $\theta = 0.7$ achieves the best balance: lower thresholds ($\theta \leq 0.6$) introduce noisy RELATED edges that crowd out critical REFERENCES and DEPENDS edges within the token budget, while higher thresholds ($\theta \geq 0.8$) miss genuinely related content. Importantly, performance is relatively stable in the range 0.6--0.8 (within 1.1\%), confirming robustness of this design choice. This is consistent with the standard practice for high-confidence semantic similarity \cite{kandola2002learning}, where $\theta = 0.7$ balances precision (avoiding spurious edges) and recall (capturing semantic relationships).

\vspace{-0.0in}
\section{Conclusion}\vspace{-0in}
We introduce \ABBR, a dependency-aware graph retrieval framework for agentic document editing that achieves 76\% consistency versus 56\% baseline while reducing token usage by 85\%. Our benchmark of 1,900 test cases enables evaluation across diverse document types and edit scenarios. Notably, \ABBR with minimal internal reasoning outperforms baselines relying on expensive extended reasoning, showing that explicit structural representations can substitute for costly computation. The  edge structure (REFERENCES, DEPENDS, CONTAINS) maps naturally to code artifacts such as call graphs and data-flow dependencies, suggesting extensions to code refactoring and software maintenance. Future directions include richer dependency detection incorporating coreference resolution and discourse structure
. Domain-adaptive learned weights for edge prioritization is a promising avenue for specialized applications.

\newpage

\section*{Limitations}
While \ABBR demonstrates consistent improvements across diverse test cases, several limitations warrant discussion. First, the framework is optimized for structured documents with clear hierarchical organization and explicit identifiers (sections, figures, equations), and may perform suboptimally on unstructured content such as creative writing, conversational text, or documents lacking consistent formatting conventions. Second, our evaluation focuses primarily on English-language documents; the dependency extraction protocols may require adaptation for languages with different syntactic structures, particularly for implicit dependency detection that relies on coreference resolution. Third, although we demonstrate generalization across six models, the quality of dependency extraction and editing operations remains bounded by the underlying LLM's capabilities—weaker models may produce less accurate dependency graphs, partially offsetting efficiency gains. Fourth, the framework assumes a relatively stable document structure during editing sessions; operations that fundamentally reorganize document architecture (e.g., merging chapters, restructuring entire sections) may require full graph reconstruction rather than incremental updates. Fifth, our experiments focus on single-agent editing scenarios, and the framework's behavior under concurrent multi-agent editing with potential conflicts requires further investigation. Sixth, while we evaluate consistency through reference validity and structural coherence metrics, these may not fully capture subtle semantic inconsistencies that human readers would detect. Finally, the constant token usage achieved by \ABBR represents an average across test cases; pathological documents with extremely dense interconnections might require retrieving larger context subgraphs, though this remains substantially more efficient than full-document processing.

\section*{Potential Risk}

The proposed method's effectiveness depends critically on the accuracy of LLM-based dependency extraction protocols. Errors in identifying explicit references, implicit dependencies, or semantic relationships during graph construction propagate through subsequent editing operations, potentially causing the system to miss critical dependencies or retrieve irrelevant context. The extraction protocols require careful prompt engineering and may exhibit inconsistent performance across document domains, with technical specifications having clearer structural patterns than creative or conversational content. Additionally, the graph construction and incremental update processes introduce computational overhead: while amortized across multiple edits, the initial graph construction for very large documents (>100,000 tokens) or documents with dense interconnections can be time-consuming. The semantic similarity threshold for RELATED edges represents a design choice that balances precision and recall but may require domain-specific tuning. Furthermore, the method assumes document structure can be reliably parsed into DOM trees, which may fail for poorly formatted documents, legacy formats, or content with inconsistent markup. Finally, the framework's reliance on external graph memory introduces a synchronization challenge: if the graph becomes desynchronized from the document state due to external edits or system failures, consistency guarantees no longer hold until the graph is reconstructed.

\section*{Ethical Considerations}
The deployment of AI agents for automated document editing raises several ethical concerns that warrant careful consideration. First, the system's ability to maintain cross-document consistency and propagate changes efficiently could facilitate unauthorized modifications or manipulation of collaborative documents if access controls are inadequate, particularly in settings involving multiple stakeholders or version-controlled content. Second, while the framework assists human editors, over-reliance on automated editing agents may reduce critical human oversight, potentially allowing errors or inappropriate modifications to propagate undetected through dependency chains. The LLM-based dependency extraction and editing components may also inherit biases present in their training data, potentially affecting how relationships are identified or how edits are formulated across different document types, domains, or writing styles. Third, the system processes document content to construct semantic graphs and generate embeddings, raising privacy concerns when applied to sensitive or confidential documents—organizations must ensure appropriate data handling practices and consider on-premises deployment for sensitive applications. Fourth, the framework's effectiveness in maintaining consistency might create false confidence in fully automated editing workflows, whereas human judgment remains essential for evaluating semantic appropriateness, stylistic coherence, and domain-specific correctness that automated systems cannot fully capture. We emphasize that \ABBR is designed as an assistive tool augmenting human editors rather than replacing them, and recommend that deployed systems include human-in-the-loop verification for critical editing operations, maintain comprehensive audit trails, and provide transparency about which modifications were AI-generated versus human-authored.

\bibliography{references} 

@article{an2024make,
  title={Make your llm fully utilize the context},
  author={An, Shengnan and Ma, Zexiong and Lin, Zeqi and Zheng, Nanning and Lou, Jian-Guang and Chen, Weizhu},
  journal={Advances in Neural Information Processing Systems},
  volume={37},
  pages={62160--62188},
  year={2024}
}

@inproceedings{laban2024summary,
  title={Summary of a haystack: A challenge to long-context llms and rag systems},
  author={Laban, Philippe and Fabbri, Alexander Richard and Xiong, Caiming and Wu, Chien-Sheng},
  booktitle={Proceedings of the 2024 Conference on Empirical Methods in Natural Language Processing},
  pages={9885--9903},
  year={2024}
}

@article{lai2025flexprefill,
  title={Flexprefill: A context-aware sparse attention mechanism for efficient long-sequence inference},
  author={Lai, Xunhao and Lu, Jianqiao and Luo, Yao and Ma, Yiyuan and Zhou, Xun},
  journal={arXiv preprint arXiv:2502.20766},
  year={2025}
}

@article{martins2020sparse,
  title={Sparse and continuous attention mechanisms},
  author={Martins, Andr{\'e} and Farinhas, Ant{\'o}nio and Treviso, Marcos and Niculae, Vlad and Aguiar, Pedro and Figueiredo, Mario},
  journal={Advances in Neural Information Processing Systems},
  volume={33},
  pages={20989--21001},
  year={2020}
}

@inproceedings{song2024low,
  title={Low-rank approximation for sparse attention in multi-modal llms},
  author={Song, Lin and Chen, Yukang and Yang, Shuai and Ding, Xiaohan and Ge, Yixiao and Chen, Ying-Cong and Shan, Ying},
  booktitle={Proceedings of the IEEE/CVF Conference on Computer Vision and Pattern Recognition},
  pages={13763--13773},
  year={2024}
}

@inproceedings{hooper2025squeezed,
  title={Squeezed attention: Accelerating long context length llm inference},
  author={Hooper, Coleman Richard Charles and Kim, Sehoon and Mohammadzadeh, Hiva and Maheswaran, Monishwaran and Zhao, Sebastian and Paik, June and Mahoney, Michael W and Keutzer, Kurt and Gholami, Amir},
  booktitle={Proceedings of the 63rd Annual Meeting of the Association for Computational Linguistics (Volume 1: Long Papers)},
  pages={32631--32652},
  year={2025}
}

@article{liu2024retrievalattention,
  title={Retrievalattention: Accelerating long-context llm inference via vector retrieval},
  author={Liu, Di and Chen, Meng and Lu, Baotong and Jiang, Huiqiang and Han, Zhenhua and Zhang, Qianxi and Chen, Qi and Zhang, Chengruidong and Ding, Bailu and Zhang, Kai and others},
  journal={arXiv preprint arXiv:2409.10516},
  year={2024}
}

@article{maharana2024evaluating,
  title={Evaluating very long-term conversational memory of llm agents},
  author={Maharana, Adyasha and Lee, Dong-Ho and Tulyakov, Sergey and Bansal, Mohit and Barbieri, Francesco and Fang, Yuwei},
  journal={arXiv preprint arXiv:2402.17753},
  year={2024}
}

@article{wang2023augmenting,
  title={Augmenting language models with long-term memory},
  author={Wang, Weizhi and Dong, Li and Cheng, Hao and Liu, Xiaodong and Yan, Xifeng and Gao, Jianfeng and Wei, Furu},
  journal={Advances in Neural Information Processing Systems},
  volume={36},
  pages={74530--74543},
  year={2023}
}

@article{chen2021multitask,
  title={Multitask offloading strategy optimization based on directed acyclic graphs for edge computing},
  author={Chen, Jiawen and Yang, Yajun and Wang, Chenyang and Zhang, Heng and Qiu, Chao and Wang, Xiaofei},
  journal={IEEE Internet of Things Journal},
  volume={9},
  number={12},
  pages={9367--9378},
  year={2021},
  publisher={IEEE}
}

@article{kandola2002learning,
  title={Learning semantic similarity},
  author={Kandola, Jaz and Cristianini, Nello and Shawe-taylor, John},
  journal={Advances in neural information processing systems},
  volume={15},
  year={2002}
}

@article{zhang2025agentic,
  title={Agentic context engineering: Evolving contexts for self-improving language models},
  author={Zhang, Qizheng and Hu, Changran and Upasani, Shubhangi and Ma, Boyuan and Hong, Fenglu and Kamanuru, Vamsidhar and Rainton, Jay and Wu, Chen and Ji, Mengmeng and Li, Hanchen and others},
  journal={arXiv preprint arXiv:2510.04618},
  year={2025}
}

@inproceedings{vaswani2017attention,
  title={Attention is All You Need},
  author={Vaswani, Ashish and Shazeer, Noam and Parmar, Niki and Uszkoreit, Jakob and Jones, Llion and Gomez, Aidan N and Kaiser, Lukasz and Polosukhin, Illia},
  booktitle={Advances in Neural Information Processing Systems},
  volume={30},
  year={2017}
}

@inproceedings{jin2024llm,
  title={{LLM} Maybe {LongLM}: Self-Extend {LLM} Context Window Without Tuning},
  author={Jin, Hongye and Han, Xiaotian and Yang, Jingfeng and Jiang, Zhimeng and Liu, Zirui and Chang, Chia-Yuan and Chen, Huiyuan and Hu, Xia},
  booktitle={International Conference on Machine Learning},
  year={2024}
}

@inproceedings{xiao2024infllm,
  title={{InfLLM}: Training-Free Long-Context Extrapolation for {LLM}s with an Efficient Context Memory},
  author={Xiao, Chaojun and Zhang, Pengle and Hu, Xu and Liu, Guangxuan and Xu, Yichuan and Lin, Chenyang and Wang, Chengqiang and Dong, Yishi and Wei, Furu and Cao, Yeyun},
  booktitle={Advances in Neural Information Processing Systems},
  volume={37},
  year={2024}
}

@inproceedings{wang2024executable,
  title={Executable Code Actions Elicit Better {LLM} Agents},
  author={Wang, Xingyao and Chen, Yangyi and Yuan, Lifan and Zhang, Yizhe and Li, Yunzhu and Peng, Hao and Ji, Heng},
  booktitle={International Conference on Machine Learning},
  year={2024}
}

@inproceedings{shinn2023reflexion,
  title={Reflexion: Language Agents with Verbal Reinforcement Learning},
  author={Shinn, Noah and Cassano, Federico and Labash, Beck and Gopinath, Ashwin and Narasimhan, Karthik and Yao, Shunyu},
  booktitle={Advances in Neural Information Processing Systems},
  volume={36},
  year={2023}
}

@inproceedings{lewis2020retrieval,
  title={Retrieval-Augmented Generation for Knowledge-Intensive {NLP} Tasks},
  author={Lewis, Patrick and Perez, Ethan and Piktus, Aleksandra and Petroni, Fabio and Karpukhin, Vladimir and Goyal, Naman and K{\"u}ttler, Heinrich and Lewis, Mike and Yih, Wen-tau and Rockt{\"a}schel, Tim and Riedel, Sebastian and Kiela, Douwe},
  booktitle={Advances in Neural Information Processing Systems},
  volume={33},
  pages={9459--9474},
  year={2020}
}

@article{liu2024lost,
  title={Lost in the Middle: How Language Models Use Long Contexts},
  author={Liu, Nelson F and Lin, Kevin and Hewitt, John and Paranjape, Ashwin and Bevilacqua, Michele and Petroni, Fabio and Liang, Percy},
  journal={Transactions of the Association for Computational Linguistics},
  volume={12},
  pages={157--173},
  year={2024},
  publisher={MIT Press}
}

@inproceedings{chen2023teaching,
  title={Teaching Large Language Models to Self-Debug},
  author={Chen, Xinyun and Lin, Maxwell and Sch{\"a}rli, Nathanael and Zhou, Denny},
  booktitle={International Conference on Learning Representations},
  year={2023}
}

@inproceedings{allamanis2018learning,
  title={Learning to Represent Programs with Graphs},
  author={Allamanis, Miltiadis and Brockschmidt, Marc and Khademi, Mahmoud},
  booktitle={International Conference on Learning Representations},
  year={2018}
}

@inproceedings{guo2021graphcodebert,
  title={{GraphCodeBERT}: Pre-training Code Representations with Data Flow},
  author={Guo, Daya and Ren, Shuo and Lu, Shuai and Feng, Zhangyin and Tang, Duyu and Liu, Shujie and Zhou, Long and Duan, Nan and Svyatkovskiy, Alexey and Fu, Shengyu and Tufano, Michele and Deng, Shao Kun and Clement, Colin and Drain, Dawn and Sundaresan, Neel and Yin, Jian and Jiang, Daxin and Zhou, Ming},
  booktitle={International Conference on Learning Representations},
  year={2021}
}

@inproceedings{yasunaga2021qa,
  title={{QA-GNN}: Reasoning with Language Models and Knowledge Graphs for Question Answering},
  author={Yasunaga, Michihiro and Ren, Hongyu and Bosselut, Antoine and Liang, Percy and Leskovec, Jure},
  booktitle={Proceedings of the 2021 Conference of the North American Chapter of the Association for Computational Linguistics: Human Language Technologies},
  pages={535--546},
  year={2021}
}

@inproceedings{zhang2022greaseLM,
  title={{GreaseLM}: Graph {REA}soning Enhanced Language Models for Question Answering},
  author={Zhang, Xikun and Bosselut, Antoine and Yasunaga, Michihiro and Ren, Hongyu and Liang, Percy and Manning, Christopher D and Leskovec, Jure},
  booktitle={International Conference on Learning Representations},
  year={2022}
}

@inproceedings{huang2022language,
  title={Language Models as Zero-Shot Planners: Extracting Actionable Knowledge for Embodied Agents},
  author={Huang, Wenlong and Abbeel, Pieter and Pathak, Deepak and Mordatch, Igor},
  booktitle={International Conference on Machine Learning},
  year={2022}
}

@inproceedings{kipf2017semi,
  title={Semi-Supervised Classification with Graph Convolutional Networks},
  author={Kipf, Thomas N and Welling, Max},
  booktitle={International Conference on Learning Representations},
  year={2017}
}

@inproceedings{hamilton2017inductive,
  title={Inductive Representation Learning on Large Graphs},
  author={Hamilton, William L and Ying, Rex and Leskovec, Jure},
  booktitle={Advances in Neural Information Processing Systems},
  volume={30},
  year={2017}
}

@inproceedings{xu2019how,
  title={How Powerful are Graph Neural Networks?},
  author={Xu, Keyulu and Hu, Weihua and Leskovec, Jure and Jegelka, Stefanie},
  booktitle={International Conference on Learning Representations},
  year={2019}
}
\newpage

\appendix

\section*{\Large Appendix}
\vspace{-0.05in}
\begin{description}[leftmargin=0.2in, itemsep=3pt, parsep=0pt]
    \item[Appendix \ref{app:prompts}] \textbf{Dependency Extraction Protocols} 
    \begin{itemize}[leftmargin=0.3in, itemsep=0pt]
        \item[$\circ$] Explicit Reference Relation ($r_{\text{exp}}$)
        \item[$\circ$] Implicit Semantic Dependency Relation ($r_{\text{imp}}$)
        \item[$\circ$] Semantic Relatedness Relation ($r_{\text{sem}}$)
        \item[$\circ$] Implementation Notes
    \end{itemize}
        
    \item[Appendix \ref{app:B}] \textbf{LEDGER Aided Document Editing Examples}
    \begin{itemize}[leftmargin=0.3in, itemsep=0pt]
        \item[$\circ$] Example Task: Revising a Theorem Statement with Cascading Updates
    \end{itemize}
    
    \item[Appendix \ref{app:evaluationmetric}] \textbf{Evaluation Metrics and Examples}
    \begin{itemize}[leftmargin=0.3in, itemsep=0pt]
        \item[$\circ$] Metric Definitions
        \item[$\circ$] Evaluation Examples (3 cases)
        \item[$\circ$] Metric Computation
    \end{itemize}
    
    \item[Appendix \ref{app:benchmark}] \textbf{Benchmark Construction}
    \begin{itemize}[leftmargin=0.3in, itemsep=0pt]
        \item[$\circ$] Benchmark Overview
        \item[$\circ$] Test Scenario Taxonomy (7 categories)
        \item[$\circ$] Document Generation Process
        \item[$\circ$] Benchmark Statistics
        \item[$\circ$] Evaluation Protocol
    \end{itemize}
    
    \item[Appendix \ref{app:additional_exp}] \textbf{Additional Experimental Results}
    \begin{itemize}[leftmargin=0.3in, itemsep=0pt]
        \item[$\circ$] Method Comparison: Context and Structural Awareness
        \item[$\circ$] Extended Baseline Comparison (including Hierarchical RAG)
        \item[$\circ$] Progressive Edge-Type Ablation
        \item[$\circ$] Retrieval Weight Sensitivity Across Document Sizes
    \end{itemize}

    \item[Appendix \ref{app:alg}] \textbf{Pseudocode for LEDGER Algorithms}
    \begin{itemize}[leftmargin=0.3in, itemsep=0pt]
        \item[$\circ$] Semantic Graph Construction
        \item[$\circ$] Dependency-Aware Context Retrieval
        \item[$\circ$] Consistency Verification
    \end{itemize}
\end{description}
\vspace{0.15in}

\newpage 
\begin{figure*}[hbt!]
    \centering
\includegraphics[width=0.94\linewidth]{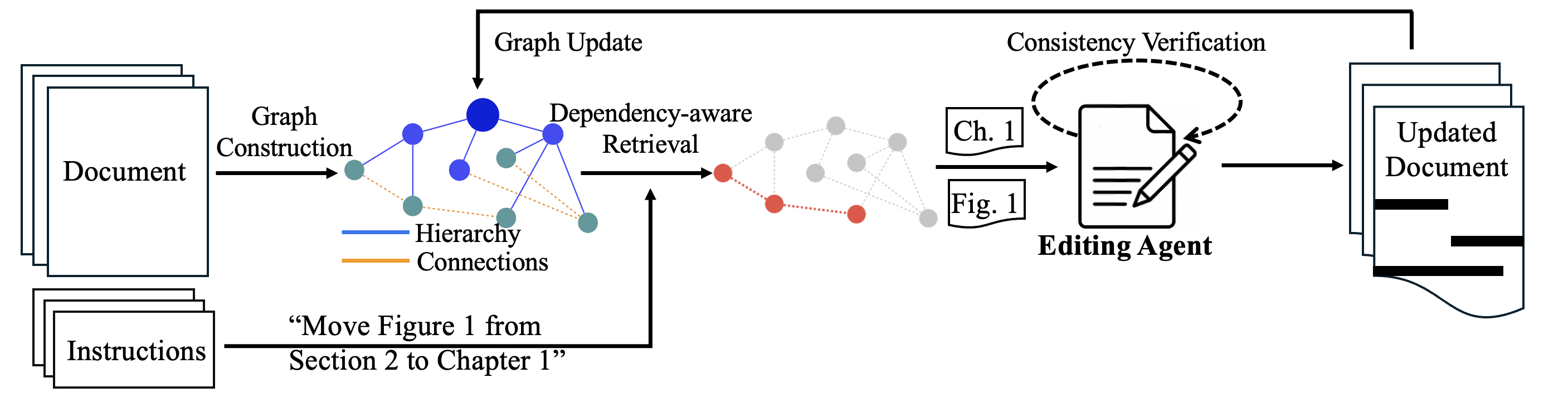}
\caption{\ABBR workflow for document editing with dependency awareness. The system constructs a semantic graph from the input document, capturing hierarchical relationships (solid blue lines) and implict, explict and semantic connections (dotted orange lines). When an edit instruction is received, the graph guides context retrieval to identify relevant dependencies (highlighted in red). The editing agent applies edits with consistency verification, and the semantic graph is updated incrementally. Example operations include updating cross-references (e.g., figure references), reorganizing sections (e.g., update Chapter 1 $\rightarrow$ 2), and adding examples while maintaining consistency across all dependent elements.}
    \label{fig:structure}
\end{figure*}

\section{Dependency Extraction Protocols}
\label{app:prompts}

This section provides the complete extraction protocols for each dependency type. Each protocol is designed to operationalize the formal relation definitions from Section~3 through structured constraints that ensure mathematical rigor.

\begin{mdframed}
   \textbf{Remark on the token cost for Graph Construction.}     Graph construction incurs a one-time cost of $O(n)$ for document parsing and embedding computation, plus $O(n\cdot k)$ for dependency extraction, where k is the average number of candidate dependencies per unit (typically 5-10 nearby units for implicit dependencies, avoiding all-pairs $O(n^2)$ computation). For a typical 10K token document with 100 units, this requires approximately 500 pairwise LLM calls for implicit dependency extraction, completing in 2-3 minutes. This cost is amortized across multiple editing operations on the same document, with incremental updates (Algorithm 2) adding minimal overhead per edit.
\end{mdframed}

\subsection{Explicit Reference Relation ($r_{\text{exp}}$)}

\textbf{Operator definition reminder:}
$r_{\text{exp}}(u_i,u_j)=1$ if and only if there exists a minimal span $m\subset u_i$ that explicitly refers to identifier $\mathrm{id}_j$.

\textbf{Extraction Protocol for $r_{\text{exp}}$:}

You are given a text unit $u_i$ from a technical document with known identifiers $\{\mathrm{id}_1, \ldots, \mathrm{id}_n\}$.

\textbf{Task:} Identify all substrings $m \subset u_i$ that establish explicit references to other units.

\textbf{Constraints:}
\begin{enumerate}
    \item \textbf{Explicit naming:} The substring $m$ must explicitly and unambiguously refer to another document unit by identifier (e.g., section number ``Section 2.1'', equation label ``Equation~3'', figure caption ``Figure~5'', named definition ``Theorem~1'').
    
    \item \textbf{Minimality:} The span $m$ must be minimal -- removing any token from $m$ must break the reference. For example, in ``as shown in Figure~3 above'', the minimal span is ``Figure~3'', not ``Figure~3 above''.
    
    \item \textbf{Surface-level only:} Do not infer references that rely on semantic similarity, paraphrasing, or topic relatedness. Only surface-level, identifier-based references are allowed. For example, ``the matrix operation discussed earlier'' is NOT an explicit reference unless it includes an identifier.
    
    \item \textbf{LaTeX references:} Include LaTeX cross-references (e.g., \verb|\ref{sec:method}|, \verb|\eqref{eq:loss}|, \verb|\cite{smith2020}|) as explicit references when the label can be resolved to $\mathrm{id}_j$.
\end{enumerate}

\textbf{Output format:} Return a set of pairs $\{(m_1, \mathrm{id}_{j_1}), (m_2, \mathrm{id}_{j_2}), \ldots\}$ where each $m_k$ is the minimal substring and $\mathrm{id}_{j_k}$ is the referenced identifier. If no such spans exist, return an empty set.

\textbf{Example:}
\begin{itemize}
    \item Input: ``The method described in Section~2.1 extends Theorem~4 from \cite{vaswani2017attention}.''
    \item Output: {(``Section~2.1'', \text{``sec-2-1''}), (``Theorem~4'', \text{``thm-4''}), (``\cite{vaswani2017attention}'', \text{``cite-vaswani2017attention''})}
\end{itemize}

\subsection{Implicit Semantic Dependency Relation ($r_{\text{imp}}$)}

\textbf{Operator definition reminder:}
$r_{\text{imp}}(u_i,u_j)=1$ if and only if $u_j$ provides semantic prerequisites necessary to interpret $u_i$.

\textbf{Extraction Protocol for $r_{\text{imp}}$:}

You are given two text units $(u_i,u_j)$ from the same document, where $u_j$ appears before $u_i$ in document order.

\textbf{Task:} Determine whether $u_j$ is a \emph{necessary semantic prerequisite} for understanding $u_i$.

\textbf{Test procedure:}
\begin{enumerate}
    \item \textbf{Counterfactual removal:} Consider $u_i$ in isolation, assuming that $u_j$ (and only $u_j$) is removed from the document. All other units remain available.
    
    \item \textbf{Semantic completeness check:} Decide whether $u_i$ becomes semantically incomplete, ambiguous, or ill-defined due to missing content from $u_j$. Specifically, check if $u_i$ relies on:
    \begin{itemize}
        \item Definitions or terminology introduced in $u_j$
        \item Mathematical notation or variables defined in $u_j$
        \item Assumptions, constraints, or problem setup from $u_j$
        \item Conceptual constructs or frameworks established in $u_j$
        \item Results, lemmas, or intermediate conclusions from $u_j$
    \end{itemize}
    
    \item \textbf{Specificity requirement:} Identify the specific elements in $u_i$ that rely on content introduced in $u_j$. The dependency must be concrete, not merely topical similarity.
    
    \item \textbf{Generic knowledge exclusion:} The dependency must not be resolvable by generic background knowledge in the domain. For example, if $u_i$ uses the term ``gradient descent'' and $u_j$ defines it, but gradient descent is common knowledge in machine learning, this does NOT establish $r_{\text{imp}}(u_i, u_j)$.
    
    \item \textbf{Directionality:} Verify that the dependency is strictly directional: $u_j \rightarrow u_i$. The relationship cannot be bidirectional or circular.
\end{enumerate}

\textbf{Output format:} Return \texttt{True} if all conditions are satisfied, establishing $r_{\text{imp}}(u_i, u_j) = 1$. Otherwise, return \texttt{False}. If \texttt{True}, also return the specific elements in $u_i$ that depend on $u_j$.

\textbf{Example:}
\begin{itemize}
    \item $u_j$: ``We define the loss function $\mathcal{L}(\theta) = \sum_{i=1}^n (y_i - f_\theta(x_i))^2$.''
    \item $u_i$: ``To minimize $\mathcal{L}(\theta)$, we compute the gradient with respect to $\theta$.''
    \item Output: \texttt{True}, with dependency elements: $\{$``$\mathcal{L}(\theta)$'', ``loss function definition''$\}$
    \item Justification: Without $u_j$, the symbol $\mathcal{L}(\theta)$ in $u_i$ is undefined.
\end{itemize}

\subsection{Semantic Relatedness Relation ($r_{\text{sem}}$)}

\textbf{Operator definition reminder:}
$r_{\text{sem}}(u_i,u_j)=1$ if and only if units $u_i$ and $u_j$ discuss semantically related topics, quantified as $\text{sim}(e_i, e_j) > \theta$ where $e_i, e_j$ are embedding vectors and $\theta$ is a threshold.

\textbf{Extraction Protocol for $r_{\text{sem}}$:}

You are given a text unit $u_i$ of type $\tau_i \in \{\text{section}, \text{paragraph}, \text{figure}, \text{table}, \text{equation}\}$.

\textbf{Task:} Generate a concise summary $s_i$ that captures the semantic content of $u_i$ for embedding-based similarity computation.

\textbf{Type-specific protocols:}

\textbf{For sections ($\tau_i = \text{section}$):}
\begin{enumerate}
    \item Extract the main claim or thesis of the section (1 sentence)
    \item Identify 2-3 key contributions, findings, or concepts introduced
    \item Exclude: Structural markers (``In this section...''), transitional phrases, citations
    \item Target length: 50-100 tokens
\end{enumerate}

\textbf{For paragraphs ($\tau_i = \text{paragraph}$):}
\begin{enumerate}
    \item Identify the central idea or main point (1-2 sentences)
    \item Include key technical terms or concepts
    \item Exclude: Supporting details, examples, citations
    \item Target length: 30-50 tokens
\end{enumerate}

\textbf{For figures ($\tau_i = \text{figure}$):}
\begin{enumerate}
    \item Use the figure caption directly as $s_i$
    \item If caption is very long ($>$100 tokens), extract the first sentence and key technical terms
    \item Include figure type (e.g., ``graph showing'', ``architecture diagram of'')
\end{enumerate}

\textbf{For tables ($\tau_i = \text{table}$):}
\begin{enumerate}
    \item Use the table caption as $s_i$
    \item If caption is uninformative, describe what is being compared or measured (e.g., ``Performance comparison of methods on datasets'')
    \item Include column/row headers if they convey key concepts
\end{enumerate}

\textbf{For equations ($\tau_i = \text{equation}$):}
\begin{enumerate}
    \item Extract any accompanying description or inline explanation
    \item If no description exists, describe the equation type (e.g., ``optimization objective'', ``probability distribution'')
    \item Target length: 20-40 tokens
\end{enumerate}

\textbf{Output format:} Return summary $s_i$ as a coherent text string. The embedding $e_i = \text{Embed}(s_i)$ is computed from this summary.

\textbf{Semantic edge creation:} After computing embeddings for all units, create edge $(v_i, v_j, \texttt{RELATED})$ when:
\begin{itemize}
    \item $\text{sim}(e_i, e_j) = \frac{e_i \cdot e_j}{\|e_i\| \|e_j\|} > \theta$ where $\theta = 0.7$
    \item $(v_i, v_j) \notin E_{\texttt{REF}} \cup E_{\texttt{DEP}}$ (not already connected by explicit or implicit dependency)
\end{itemize}

\textbf{Example:}
\begin{itemize}
    \item Input (section): ``This section introduces our novel attention mechanism. We propose a multi-head cross-attention layer that attends to both encoder and decoder states. Experiments show significant improvements over baseline transformers.''
    \item Output: ``Novel attention mechanism using multi-head cross-attention over encoder and decoder states, improving transformer performance.'' (18 tokens)
\end{itemize}

\subsection{Implementation Notes}

\textbf{Validation and quality control:}
\begin{itemize}
    \item For $r_{\text{exp}}$: Validate that all extracted $\mathrm{id}_j$ exist in the known identifier set $\{\mathrm{id}_1, \ldots, \mathrm{id}_n\}$. Discard any references to non-existent identifiers.
    
    \item For $r_{\text{imp}}$: Apply the protocol to unit pairs $(u_i, u_j)$ where $j < i$ (i.e., $u_j$ precedes $u_i$). The test is computationally expensive, so prioritize pairs where $u_i$ is close to $u_j$ in document order (e.g., within 5-10 units).
    
    \item For $r_{\text{sem}}$: Compute embeddings using a consistent model (e.g., \texttt{text-embedding-ada-002}, \texttt{sentence-transformers/all-mpnet-base-v2}). The threshold $\theta = 0.7$ balances precision (avoiding spurious edges) and recall (capturing true semantic relationships).
\end{itemize}

\textbf{Edge priority:} When multiple edge types could connect the same unit pair, prioritize: $E_{\texttt{REF}} > E_{\texttt{DEP}} > E_{\texttt{REL}}$. This ensures explicit dependencies take precedence over semantic relationships.

\section{\ABBR aided Document Editing Examples}
\label{app:B}

\subsection{Example Task: Revising a Theorem Statement with Cascading Updates}

\textbf{Task Description:} Revise Theorem 2 in Section 3.2 by tightening the revenue bound from $O(\log n)$ to $O(1)$, ensuring all dependent sections remain consistent with the updated statement.

\textbf{Document Context:} A 10,000-token academic paper on dynamic pricing for hotel room allocation, containing 7 sections with multiple cross-references between theoretical results, algorithm descriptions, case study validation, and discussion.

\subsubsection{Dependency Graph Construction}

\ABBR constructs a semantic graph identifying three types of dependencies:

\paragraph{Explicit References:}
\begin{itemize}
    \item Theorem 2 (Section 3.2) $\leftarrow$ cited by [Section 4.para1, Section 5.para2, Section 6.para1]
    \item Lemma 1 (Section 3.1) $\leftarrow$ cited by [Section 3.2.para2, Section 4.para2]
    \item Equation (5) (Section 2.2) $\leftarrow$ cited by [Section 3.1.proof, Section 4.para3, Section 5.para3]
\end{itemize}

\paragraph{Implicit References:}
\begin{itemize}
    \item ``the optimality guarantee'' (Section 4.para1) $\rightarrow$ Theorem 2
    \item ``our main theorem'' (Section 5.para2) $\rightarrow$ Theorem 2
    \item ``this bound'' (Section 3.2.proof.para2) $\rightarrow$ Lemma 1
\end{itemize}

\paragraph{Semantic Dependencies:}
\begin{itemize}
    \item Section 4.para1 \texttt{builds\_on} Theorem 2 (algorithmic extension)
    \item Section 5.para2 \texttt{validates} Theorem 2 (empirical confirmation)
    \item Section 6.para1 \texttt{extends} Theorem 2 (multi-hotel generalization)
\end{itemize}

\subsubsection{Edit Execution with Dependency Tracking}

\ABBR identifies that modifying Theorem 2 triggers the critical dependency rule: \textit{If Theorem 2 changes $\rightarrow$ MUST verify [4.para1, 5.para2, 6.para1]}. The agent executes targeted edits in dependency order:

\paragraph{Step 1: Update Theorem 2 statement (Section 3.2)}
\begin{quote}
\textbf{Original:} ``...within $O(\log n)$ of the optimal offline benchmark''\\
\textbf{Revised:} ``...within $O(1)$ of the optimal offline benchmark''
\end{quote}

\paragraph{Step 2: Update proof outline to align with new bound}
\begin{quote}
\textbf{Original:} ``Combining these bounds across $n$ periods and applying harmonic series properties yields the $O(\log n)$ gap''\\
\textbf{Revised:} ``...we charge each deviation to a constant decrease in a suitable potential function, implying only $O(1)$ total loss''
\end{quote}

\paragraph{Steps 3--7: Propagate to dependent sections}
\begin{itemize}
    \item Section 3.2 proof paragraph: Update to ``horizon-independent constant bound''
    \item Section 4.para1: Change ``optimality guarantee'' to ``constant-gap optimality guarantee''
    \item Section 4.para3: Update reference to ``constant-gap guarantee specified in Theorem 2''
    \item Section 5.para2: Replace ``$O(\log n)$ bound'' with ``$O(1)$ bound''
    \item Section 6.para1: Insert ``preserving the constant additive gap''
\end{itemize}

\subsubsection{Efficiency Analysis}

\paragraph{Token Utilization:}
\begin{itemize}
    \item Total document: 10,000 tokens
    \item Retrieved context: $\sim$1,200 tokens (12\% of document)
    \begin{itemize}
        \item Theorem 2 statement and proof
        \item Explicit dependencies: [4.para1, 4.para3, 5.para2, 6.para1]
        \item Implicit reference text
    \end{itemize}
    \item Processing overhead: 7 targeted paragraph replacements
\end{itemize}

\paragraph{Comparison to Baseline Approaches:}
\begin{itemize}
    \item \textbf{Full document rewrite:} Would require processing all 10,000 tokens, with risk of introducing inconsistencies in unrelated sections
    \item \textbf{Local edit without dependency tracking:} Would update Theorem 2 but miss the 6 dependent references, breaking document consistency
    \item \textbf{\ABBR approach:} Processes 12\% of document tokens while maintaining global consistency through explicit dependency tracking
\end{itemize}

\subsubsection{Consistency Verification}

\ABBR ensures consistency across multiple dimensions:

\begin{enumerate}
    \item \textbf{Referential Integrity:} All explicit citations to Theorem 2 now correctly describe the $O(1)$ bound
    \item \textbf{Semantic Coherence:} Algorithm design and case study sections align with the tightened theoretical guarantee
    \item \textbf{Proof Validity:} Proof outline mechanism (potential-based charging) supports the stronger constant bound
    \item \textbf{Terminology Consistency:} Uniform use of ``constant additive gap'' and ``$O(1)$ bound'' throughout dependent sections
\end{enumerate}

The edit preserves document flow while implementing a substantial theoretical improvement that cascades through algorithm description, empirical validation, and discussion sections---demonstrating \ABBR's capability to maintain consistency in complex, interconnected technical documents.

\section{Evaluation Metrics and Examples}
\label{app:evaluationmetric}

This appendix provides formal definitions of the evaluation metrics used to assess \ABBR performance, along with concrete examples from our benchmark dataset illustrating how each metric is evaluated.

\subsection{Metric Definitions}

We evaluate system performance across seven complementary metrics organized into three categories: dependency awareness, efficiency, and quality.

\textbf{Dependency Awareness Metrics:}

\begin{itemize}
\item \textbf{Dependency Detection (Explicit):} Measures whether the system correctly identifies all explicit references in the document, including section citations (``Section 3.2''), figure references (``Figure 1''), table references (``Table 2''), equation citations (``Equation (1)''), and algorithm references (``Algorithm 1''). This metric is computed as the percentage of explicit references correctly identified and included in the retrieved context.

\item \textbf{Dependency Detection (Implicit):} Measures whether the system correctly resolves implicit references such as anaphoric expressions (``this approach'', ``the metric'', ``these results'') and contextual references (``the preceding computation'', ``as described earlier''). This metric is computed as the percentage of implicit references correctly resolved to their referents.

\item \textbf{Dependency Detection (Semantic):} Measures whether the system recognizes semantic relationships between document elements that lack explicit citations, such as conceptual dependencies (one section building on another), comparisons (contrasting methods), and synthesis (combining results). This metric is computed as the percentage of semantic dependencies correctly identified through graph edges.
\end{itemize}

\textbf{Efficiency Metrics:}

\begin{itemize}
\item \textbf{Token Efficiency:} Measures the ratio of tokens retrieved for an edit operation to the total document size. An efficient system retrieves only the minimal context necessary for consistency. This metric is computed as: $\text{Token Efficiency} = \frac{\text{Tokens Retrieved}}{\text{Total Document Tokens}} \times 100\%$. Lower percentages indicate higher efficiency, with typical values ranging from 10 to 15\% for \ABBR versus 100\% for full document approaches.
\end{itemize}

\textbf{Quality Metrics:}

\begin{itemize}
\item \textbf{Reference Validity:} Measures whether all explicit references (section numbers, figure citations, table references) remain correct after an edit operation. This metric is computed as the percentage of explicit references that correctly resolve to their intended targets after modifications. A broken reference (e.g., citing ``Figure 3'' when only Figures 1 and 2 exist) constitutes a failure.

\item \textbf{Consistency Preservation:} Measures whether edits maintain semantic coherence with dependent content. This includes verifying that implicit references remain meaningful, terminology changes propagate to dependent sections, and semantic relationships remain valid. This metric is computed as the percentage of dependent elements that remain consistent with the edited content.

\item \textbf{Edit Quality:} Measures whether the edit correctly addresses the instruction while maintaining appropriate style, clarity, and factual accuracy. This metric is computed through comparison with gold standard human edits, assessing whether the modification achieves the stated goal without introducing errors or unintended changes.
\end{itemize}

\subsection{Evaluation Examples}

We present three representative examples from our benchmark dataset to illustrate how these metrics are applied in practice.

\subsubsection{Example 1: Local Paragraph Refinement}

\begin{mdframed}[backgroundcolor=gray!5, linewidth=1pt]
\textbf{Scenario:} Academic paper (12K tokens) with 6 sections. Edit instruction: ``Improve the clarity of the second paragraph in Section 4.1 (Limitations) by presenting a more balanced discussion of memory overhead and suggesting potential mitigation strategies.''

\textbf{Document Structure:}
\begin{itemize}
\item Explicit references: Figure 2 cited in 4 locations; Definition ``token efficiency'' cited in 2 locations; Algorithm 1 cited in 3 locations
\item Implicit references: ``the metric'' in Section 4.1 refers to Definition ``token efficiency''
\item Semantic dependencies: Section 4.1 builds on Algorithm 1; Section 5 synthesizes findings from Section 4
\end{itemize}
\end{mdframed}

\textbf{Metric Evaluation:}

\textit{Dependency Detection (Explicit):} System must identify that Figure 2 appears in paragraphs 1.3, 3.2.2, 4.1, and 5.2; Algorithm 1 appears in sections 2.2 and 4.1; Definition ``token efficiency'' appears in sections 3.1 and 4.1. \textbf{Pass criterion:} All 5 explicit reference types correctly identified.

\textit{Dependency Detection (Implicit):} System must resolve that ``the metric'' in Section 4.1 paragraph 2 refers to Definition ``token efficiency'' in Section 3.1, requiring coreference resolution across sections. \textbf{Pass criterion:} Implicit reference correctly resolved and included in context.

\textit{Dependency Detection (Semantic):} System must recognize that Section 4 paragraph 1 semantically compares baseline methods with the Local Refinement Module from Section 3.2, despite no explicit citation. \textbf{Pass criterion:} Semantic comparison relationship detected.

\textit{Token Efficiency:} System should retrieve: Section 4.1 second paragraph (target, 150 tokens) + Definition ``token efficiency'' (80 tokens) + Algorithm 1 (200 tokens) + Figure 2 caption (70 tokens) + Table 1 caption (50 tokens) = approximately 550 tokens. \textbf{Pass criterion:} Retrieved tokens $\leq$ 1,500 tokens (12.5\% of 12K document).

\textit{Reference Validity:} After refinement, all explicit references in Section 4.1 paragraph 2 must correctly resolve: Algorithm 1 still exists, Definition ``token efficiency'' still exists, Section 3.1 still exists. \textbf{Pass criterion:} All explicit references remain valid.

\textit{Consistency Preservation:} Updated paragraph must maintain coherence with the implicit reference ``the metric'' and align with the memory overhead discussion in Section 5 paragraph 2. \textbf{Pass criterion:} No semantic contradictions introduced.

\textit{Edit Quality:} Paragraph provides balanced discussion of memory overhead, adds at least one concrete mitigation strategy (e.g., ``implementing lazy loading of intermediate matrices''), and improves clarity without altering factual content. \textbf{Pass criterion:} Instruction requirements satisfied, no factual errors.

\subsubsection{Example 2: Content Addition with Dependencies}

\begin{mdframed}[backgroundcolor=gray!5, linewidth=1pt]
\textbf{Scenario:} Technical document (5K tokens) explaining credit risk calculations. Edit instruction: ``Insert a concrete example illustrating the Probability of Default (PD) component of the Expected Credit Loss calculation at the end of the second paragraph of Section 3.1.''

\textbf{Document Structure:}
\begin{itemize}
\item Explicit references: Figure 2 cited in Overview, Section 3.1, Section 4.1, Section 6; Equation (1) in Section 3.1
\item Implicit references: ``the preceding computation'' in Section 3.2 refers to Equation (1)
\item Semantic dependencies: Section 6 compares Section 3 methodology with legacy approaches
\end{itemize}
\end{mdframed}

\textbf{Metric Evaluation:}

\textit{Dependency Detection (Explicit):} System identifies explicit references to Figure 2 in 4 locations, Table 1 in 3 locations, Equation (1) in Section 3.1, and Definition PD in Section 2.2. \textbf{Pass criterion:} All explicit references in relevant sections identified.

\textit{Dependency Detection (Implicit):} System resolves that ``the preceding computation'' in Section 3.2 paragraph 2 refers to Equation (1) in Section 3.1, ensuring the new example aligns with this implicit reference. \textbf{Pass criterion:} Implicit reference correctly resolved.

\textit{Dependency Detection (Semantic):} System recognizes that Section 6 paragraph 2 semantically compares Section 3 methodology with legacy spreadsheet approaches, despite no explicit citations. The new example may affect this comparison. \textbf{Pass criterion:} Semantic comparison relationship detected.

\textit{Token Efficiency:} System should retrieve: Section 3.1 all paragraphs (300 tokens) + Definition PD from Section 2.2 (50 tokens) + Table 1 (60 tokens) + Equation (1) (40 tokens) + Figure 2 caption (50 tokens) = approximately 500 tokens. \textbf{Pass criterion:} Retrieved tokens $<$ 10\% of 5K document (500 tokens).

\textit{Reference Validity:} All explicit references (Figure 2, Table 1, Equation (1), Definition PD, Section 2.2) must remain valid after inserting the new example. No renumbering should be triggered. \textbf{Pass criterion:} All explicit references still resolve correctly.

\textit{Consistency Preservation:} The implicit reference ``the preceding computation'' in Section 3.2 and semantic comparisons in Section 6 remain accurate after example insertion. The new content must not contradict existing methodology descriptions. \textbf{Pass criterion:} No contradictions with implicit or semantic dependencies.

\textit{Edit Quality:} The inserted example clearly demonstrates PD calculation using a simple numerical scenario (e.g., ``For a borrower with credit score 650 and debt ratio 0.4, the estimated PD is 2.3\%''), is positioned correctly after Section 3.1 paragraph 2, and uses consistent terminology. \textbf{Pass criterion:} Example is clear, concrete, and properly positioned.

\subsubsection{Example 3: Cross-Reference Update}

\begin{mdframed}[backgroundcolor=gray!5, linewidth=1pt]
\textbf{Scenario:} API documentation (20K tokens) with code examples. Edit instruction: ``Update the signature of the reserveRoom endpoint in Section 3.2 from reserveRoom(hotelId, guestInfo) to reserveRoom(hotel\_id, guest\_profile) and make sure every code example and reference throughout the document is updated consistently.''

\textbf{Document Structure:}
\begin{itemize}
\item Explicit references: Figure 1 cited in 3 locations; Endpoint ``/api/v1/reserveRoom'' in 3 code listings; Listings 2 and 3 cited in 2 locations
\item Implicit references: ``this endpoint'' in Section 4.1 refers to Section 3.2; ``the previous signature'' in Section 6 refers to old parameter names
\item Semantic dependencies: Listing 3 builds on authentication flow; Section 5 compares reserveRoom vs cancelBooking
\end{itemize}
\end{mdframed}

\textbf{Metric Evaluation:}

\textit{Dependency Detection (Explicit):} System identifies all explicit references: Figure 1 in sections 1.2, 3.1, and 4.1; Endpoint ``/api/v1/reserveRoom'' in Listings 2, 3, and Section 5.1; Table 1 in sections 2.2, 4.1.3, and 4.2.4. \textbf{Pass criterion:} All explicit references to endpoint and parameters identified.

\textit{Dependency Detection (Implicit):} System resolves: ``this endpoint'' in Section 4.1 paragraph 2 refers to Section 3.2 reserveRoom definition; ``these examples'' in Section 5 paragraph 2 refers to Listings 2 and 3; ``the previous signature'' in Section 6 paragraph 1 refers to reserveRoom(hotelId, guestInfo). \textbf{Pass criterion:} All implicit references correctly resolved.

\textit{Dependency Detection (Semantic):} System infers: Listing 3 builds on authentication flow from Section 4.2; Section 5 paragraph 3 compares reserveRoom versus cancelBooking semantically; Section 4.1 paragraph 3 synthesizes Table 1 with Listing 2 output. \textbf{Pass criterion:} Semantic dependencies identified.

\textit{Token Efficiency:} System should retrieve: Section 3.2 endpoint definition (200 tokens) + Listings 2 and 3 (400 tokens each) + all paragraphs referencing endpoint or parameters (600 tokens) + changelog line (50 tokens) = approximately 1,650 tokens. \textbf{Pass criterion:} Retrieved tokens $<$ 10\% of 20K document (2,000 tokens).

\textit{Reference Validity:} After updating parameter names from camelCase to snake\_case, all explicit references must still resolve correctly: Figure 1 references unchanged, endpoint path unchanged, listing citations unchanged. \textbf{Pass criterion:} All explicit references remain valid.

\textit{Consistency Preservation:} Implicit references (``this endpoint'', ``the previous signature'') and semantic comparisons remain meaningful after parameter name changes. Text referring to ``hotelId'' must be updated to ``hotel\_id'' throughout. \textbf{Pass criterion:} Terminology changes propagated to all dependent text.

\textit{Edit Quality:} reserveRoom signature updated in definition and all code examples (Listings 2, 3, and inline examples); code compiles without errors; explanatory text aligns with new parameter names using snake\_case convention consistently. \textbf{Pass criterion:} All occurrences updated, code valid, terminology consistent.

\subsection{Metric Computation}

For each test case, we compute metrics through automated validation:

\textbf{Dependency Detection Metrics:} We programmatically compare the set of dependencies identified by the system against the ground truth dependency annotations in each test case. For explicit references, we extract all citations using pattern matching (e.g., ``Figure~\textbackslash d+'', ``Section~\textbackslash d+.\textbackslash d+''). For implicit references, we verify that anaphoric expressions resolve to the correct antecedents through coreference chains. For semantic dependencies, we check that the graph contains edges representing conceptual relationships marked in the ground truth.

\textbf{Token Efficiency:} We measure the total number of tokens in the context provided to the model for each edit operation and compute the ratio to total document size. This is measured directly through tokenization of the retrieved content.

\textbf{Reference Validity:} We parse the edited document to extract all explicit references and verify that each reference target exists in the updated document. For example, a citation to ``Figure 3'' must correspond to an actual figure with that identifier. References are validated through programmatic document structure analysis.

\textbf{Consistency Preservation:} We check that implicit references remain resolvable and that terminology changes propagate correctly. For example, if a definition is modified, we verify that dependent sections using that term are also updated. We also check semantic coherence by comparing embeddings of related sections to ensure similarity remains above threshold $\theta = 0.7$.

\textbf{Edit Quality:} For a subset of 200 test cases, we compare system outputs against gold standard human edits. We assess whether the modification achieves the instruction goal, maintains appropriate style and tone, preserves factual accuracy, and introduces no unintended changes. Quality scores are aggregated across these dimensions.

Final scores for each metric represent the percentage of test cases that meet the pass criterion, averaged across all applicable test cases and models.

\section{Benchmark Construction}
\label{app:benchmark}

We construct a comprehensive benchmark to evaluate consistency preservation, efficiency, and scalability in agentic document editing. This appendix describes the test scenario taxonomy, document generation process, and evaluation framework.

\subsection{Benchmark Overview}

Our benchmark comprises 1,900 test cases spanning seven categories of editing operations, designed to systematically evaluate agent performance across diverse document structures, edit complexities, and document sizes. Each test case consists of: (1) a source document with explicit structure and dependencies, (2) a natural language editing instruction, (3) ground-truth expected modifications, and (4) validation criteria for consistency, quality, and efficiency.

The benchmark is designed around the core challenges of document editing identified in Section~1: maintaining cross-references during modifications, preserving semantic consistency across dependent sections, and scaling efficiently with document size. Test cases range from localized refinements with minimal dependencies to structural reorganizations requiring global updates, enabling fine-grained analysis of system capabilities.

\subsection{Test Scenario Taxonomy}

We organize test scenarios into seven categories, each targeting specific aspects of dependency-aware editing.

\subsubsection{Category 1: Local Refinement Edits}

Local refinements modify specific paragraphs or sections with minimal cross-document dependencies. These scenarios evaluate whether systems can apply targeted improvements without unnecessary context retrieval or unintended modifications to unrelated content.

\noindent\textbf{Subcategories:}
\begin{itemize}
    \item \textbf{Paragraph and section refinement (120 cases):} Instructions targeting clarity improvements, argument strengthening, or content simplification within localized scope. Examples include ``Improve the clarity of the second paragraph in Section 3.2'' and ``Make the proof in Section 4.1 more concise while preserving logical steps.'' These test the system's ability to retrieve minimal context (target paragraph plus containing section) while avoiding unnecessary full-document processing.
    
    \item \textbf{Local content enhancement (80 cases):} Adding examples, definitions, or explanatory text within sections. Examples include ``Add a concrete example demonstrating the algorithm after its description in Section 3.1'' and ``Insert a worked example illustrating the calculation process.'' These test whether systems can identify the appropriate insertion point and maintain local coherence without disrupting document structure.
    
    \item \textbf{Error correction (60 cases):} Fixing typos, incorrect values, broken references, or outdated information. Examples include ``Fix the typo in the third paragraph of Section 2.1'' and ``Correct the incorrect parameter value in the code example.'' These test precision in locating and modifying specific elements without introducing unintended changes.
\end{itemize}

\noindent\textbf{Expected behavior:} Systems should retrieve only the target section and immediate structural context, typically 200-800 tokens. Consistency requirements are minimal since dependencies are local. Token efficiency is the primary evaluation focus.

\subsubsection{Category 2: Cross-Reference Dependent Edits}

Cross-reference edits modify content that is referenced by or depends on multiple other sections, requiring the system to identify and update all dependent locations.

\noindent\textbf{Subcategories:}
\begin{itemize}
    \item \textbf{Figure and table caption updates (80 cases):} Modifying captions for elements referenced throughout the document. Example: ``Update Figure 3's caption to clarify the x-axis label'' where Figure 3 appears in 4-6 sections. Tests whether the system identifies all referencing sections and includes them in context to verify descriptions remain accurate.
    
    \item \textbf{Definition and terminology changes (70 cases):} Revising definitions or changing terminology used consistently across the document. Example: ``Revise the definition of 'token efficiency' in Section 2.1 to add precision'' where the term appears in 5-8 locations. Tests both explicit reference tracking (e.g., ``see Section 2.1 for the definition'') and implicit usage (paragraphs relying on the definition without citation).
    
    \item \textbf{API and code example updates (60 cases):} Changing function signatures, parameter names, or code structures used in multiple examples. Example: ``Update the reserveRoom endpoint signature from reserveRoom(hotelId, guestInfo) to reserveRoom(hotel\_id, guest\_profile) throughout all code examples.'' Tests tracking of both explicit code references and implicit dependencies where examples build on earlier code.
    
    \item \textbf{Theorem and mathematical content updates (50 cases):} Modifying theorems, lemmas, or equations referenced in proofs or derivations. Example: ``Revise Theorem 2's bound from O(n log n) to O(n) and verify dependent proofs remain valid.'' Tests mathematical dependency tracking and semantic coherence validation.
    
    \item \textbf{Citation and bibliography updates (40 cases):} Modifying citation formats, adding/removing references, or updating bibliography entries. Example: ``Change all citations to use APA format instead of ACM format.'' Tests systematic updates across the document while maintaining reference integrity.
\end{itemize}

\noindent\textbf{Expected behavior:} Systems should identify all sections containing explicit references (via \texttt{REFERENCES} edges), implicit dependencies (via \texttt{DEPENDS} edges), and semantically related content (via \texttt{RELATED} edges). Context retrieval should be proportional to the number of dependencies (typically 800-2,000 tokens for elements with 4-6 references). Consistency preservation is the primary evaluation focus.

\subsubsection{Category 3: Structural Reorganization}

Structural reorganizations involve section reordering, merging, splitting, or hierarchy changes that require updating all cross-references and maintaining document coherence.

\noindent\textbf{Subcategories:}
\begin{itemize}
    \item \textbf{Section reordering (60 cases):} Moving sections to different positions while updating numbering and cross-references. Example: ``Move Section 5 to appear before Section 3'' requires updating all references like ``as described in Section 5'' and renumbering subsequent sections. Tests whether systems track both forward and backward references.
    
    \item \textbf{Section merging and splitting (50 cases):} Combining or dividing sections while maintaining hierarchy. Example: ``Split Section 4 into two sections: 'Methodology Overview' and 'Implementation Details''' requires updating section numbers, redistributing subsections, and adjusting cross-references. Tests hierarchical consistency maintenance.
    
    \item \textbf{Section deletion and addition (30 cases):} Removing or inserting sections while updating references. Example: ``Delete Section 3.2 and merge its content into Section 3.1'' requires identifying all references to 3.2, updating them to reference 3.1, and renumbering subsequent subsections. Tests dangling reference detection.
    
    \item \textbf{Hierarchy restructuring (20 cases):} Changing nesting levels and parent-child relationships. Example: ``Promote all subsections of Section 4 to be top-level sections'' requires restructuring the containment hierarchy and updating all hierarchical references. Tests \texttt{CONTAINS} edge maintenance.
\end{itemize}

\noindent\textbf{Expected behavior:} Systems should retrieve the entire structural hierarchy affected by the reorganization, including all sections with references to moved/modified sections. Token usage varies significantly (1,500-4,000 tokens) based on reorganization scope. Both structural integrity and reference validity are critical evaluation dimensions.

\subsubsection{Category 4: Content Addition with Integration}

Content addition scenarios require inserting new material that integrates coherently with existing structure, maintaining appropriate cross-references and dependencies.

\noindent\textbf{Subcategories:}
\begin{itemize}
    \item \textbf{Example and illustration addition (60 cases):} Inserting examples, case studies, or visual elements that reference existing content. Example: ``Add a detailed walkthrough example demonstrating the three-phase workflow described in Section 2'' requires understanding the workflow definition and creating consistent terminology and structure.
    
    \item \textbf{New subsection addition (40 cases):} Adding complete subsections that extend existing sections. Example: ``Insert a new subsection comparing the proposed approach with the baseline method after Section 4.1'' requires understanding both approaches and maintaining consistent comparison criteria.
    
    \item \textbf{Cross-cutting content addition (20 cases):} Adding content that synthesizes or references multiple existing sections. Example: ``Add a paragraph synthesizing findings from Sections 3, 4, and 5 at the beginning of Section 6'' requires retrieving context from all referenced sections and maintaining semantic coherence.
\end{itemize}

\noindent\textbf{Expected behavior:} Systems should retrieve all sections referenced in the instruction plus semantically related content to ensure stylistic and terminological consistency. Context retrieval should scale with the number of sections involved (800-1,800 tokens). Coherence with existing content is the primary evaluation focus.

\subsubsection{Category 5: Multi-Round Iterative Editing}

Iterative editing scenarios simulate realistic collaborative writing workflows with sequences of 3-5 dependent edits, where each edit builds on previous modifications.

\noindent\textbf{Subcategories:}
\begin{itemize}
    \item \textbf{Sequential refinement workflow (40 cases):} Progressive refinements to improve content quality. Example: Three-step sequence: (1) ``Revise Section 3.2 to emphasize the key contribution,'' (2) ``Add supporting experimental evidence after the revised paragraph,'' (3) ``Strengthen connections between Section 3.2 and the results in Section 5.'' Tests whether the graph updates incrementally and subsequent edits have access to previous modifications.
    
    \item \textbf{Iterative restructuring workflow (35 cases):} Multiple structural changes building on each other. Example: Three-step sequence: (1) ``Move Section 4 earlier,'' (2) ``Merge it with Section 2,'' (3) ``Add new content to the merged section.'' Tests incremental graph updates and consistency maintenance across multiple structural operations.
    
    \item \textbf{Content development workflow (35 cases):} Incremental content addition across sections. Example: Three-step sequence: (1) ``Add a definition of 'context efficiency' in Section 2,'' (2) ``Add an example using this definition in Section 3,'' (3) ``Reference both the definition and example when discussing results in Section 5.'' Tests dependency creation and tracking across editing rounds.
    
    \item \textbf{Complete revision workflow (50 cases):} End-to-end revision cycles combining multiple edit types. Example: Four-step sequence: (1) ``Revise the introduction,'' (2) ``Update Figure 2's caption,'' (3) ``Strengthen the argument in Section 4,'' (4) ``Reorganize section order.'' Tests sustained consistency over complex editing workflows.
\end{itemize}

\noindent\textbf{Expected behavior:} Systems must update the dependency graph after each edit and use the updated graph for subsequent operations. Token usage should remain constant per edit even as document state evolves. Overall consistency across the editing sequence is the primary evaluation focus.

\subsubsection{Category 6: Scalability Test Scenarios}

Scalability scenarios apply identical editing operations to documents of varying lengths (5K, 10K, 20K, 50K, 100K tokens) to test whether token usage remains constant as document size increases.

\noindent\textbf{Test configurations:}
\begin{itemize}
    \item \textbf{Local edits across sizes (20 operations × 5 sizes = 100 cases):} Simple refinements like ``Improve clarity of a paragraph in Section 3.2.'' Expected token usage should remain constant (~500 tokens) regardless of document size.
    
    \item \textbf{Cross-reference edits across sizes (10 operations × 5 sizes = 50 cases):} Dependency-heavy edits like ``Update Figure 3's caption'' where the figure is referenced in 4-6 sections (constant across document sizes). Expected token usage should remain constant (~1,500 tokens) even though total document size varies by 20×.
    
    \item \textbf{Structural edits across sizes (10 operations × 5 sizes = 50 cases):} Reorganizations like ``Move Section 5 to appear before Section 3.'' Expected token usage should scale sublinearly, retrieving only affected sections rather than the entire document.
\end{itemize}

\noindent\textbf{Expected behavior:} Token usage should remain constant or scale sublinearly with document size, demonstrating $O(1)$ rather than $O(|D|)$ behavior. This validates the core efficiency claim that dependency-aware retrieval enables scalable editing.

\subsubsection{Category 7: Consistency Critical Scenarios}

Consistency-critical scenarios are designed as pass/fail tests where any consistency violation (broken reference, terminology mismatch, semantic contradiction) constitutes failure.

\noindent\textbf{Subcategories:}
\begin{itemize}
    \item \textbf{Cascading reference updates (50 cases):} Edits that should trigger automatic updates in all dependent locations. Example: ``Rename Section 4 to 'Experimental Methodology' '' should update all references like ``see Section 4'' or ``as described in the Methods section.'' Failure modes include: missing references, partially updated references, or broken cross-reference links.
    
    \item \textbf{Dependency chain edits (30 cases):} Modifications to foundational content that affects derived content. Example: ``Modify the definition of 'retrieval precision' in Section 2 to include a threshold parameter'' should propagate to all calculations, examples, and results using this metric. Failure modes include: orphaned dependencies, inconsistent terminology, or contradictory statements.
    
    \item \textbf{Structural integrity tests (40 cases):} Reorganizations that could violate hierarchy if not handled correctly. Example: ``Delete Section 3 which contains subsections 3.1, 3.2, 3.3'' should properly handle the child sections (either delete them or promote them). Failure modes include: orphaned subsections, broken containment relationships, or incorrect section numbering.
\end{itemize}

\noindent\textbf{Expected behavior:} Systems must achieve 100\% consistency on these test cases. Any broken reference, terminology inconsistency, or structural violation is considered a failure. These cases validate the reliability of dependency tracking and consistency verification mechanisms.

\subsection{Document Generation Process}

We generate synthetic documents with controlled structure and dependencies to enable systematic evaluation.

\noindent\textbf{Document structure:} Each document follows a hierarchical structure with 5-12 top-level sections, 2-5 subsections per section, and 3-8 paragraphs per subsection. Documents include structural elements (sections, paragraphs), referenced elements (figures, tables, equations), and semantic content (definitions, examples, proofs). We ensure controlled dependency density: each figure/table is referenced by 4-6 paragraphs, each definition is used in 5-8 locations, and each theorem is cited by 2-4 dependent results.

\noindent\textbf{Size variations:} We generate five document size categories by varying content length while maintaining consistent structural complexity:
\begin{itemize}
    \item \textbf{Small (5K tokens):} 5 sections, 15 subsections, 60 paragraphs
    \item \textbf{Medium (10K tokens):} 7 sections, 21 subsections, 100 paragraphs
    \item \textbf{Large (20K tokens):} 9 sections, 30 subsections, 150 paragraphs
    \item \textbf{Very large (50K tokens):} 11 sections, 40 subsections, 250 paragraphs
    \item \textbf{Extreme (100K tokens):} 12 sections, 50 subsections, 400 paragraphs
\end{itemize}

\noindent\textbf{Dependency injection:} We programmatically inject dependencies during document generation to ensure reproducibility. Explicit references are inserted using templates (``as shown in Figure X'', ``see Section Y''). Implicit dependencies are created through pronoun chains and contextual references (``This approach...'', ``The aforementioned method...''). Semantic relationships emerge from topical coherence within sections and subsections discussing related concepts.

\noindent\textbf{Ground truth:} For each test case, we generate ground-truth expected modifications by applying the edit operation programmatically to the document structure. For reference-updating edits, we identify all affected locations through dependency analysis. For structural reorganizations, we recompute section numbering and update all cross-references. For content additions, we specify insertion points and expected integration patterns. This ground truth enables automated evaluation metrics described below.

\subsection{Benchmark Statistics}

Table~\ref{tab:benchmark_stats} summarizes the distribution of test cases across categories and complexity levels.

\begin{table}[h]
\centering
\small
\begin{tabular}{@{}lrrr@{}}
\toprule
\textbf{Category} & \textbf{Cases} & \textbf{Avg. Deps} & \textbf{Avg. Tokens} \\
\midrule
Local Refinement & 260 & 1.2 & 650 \\
Cross-Reference & 300 & 5.8 & 1,520 \\
Structural Reorg. & 160 & 8.4 & 2,350 \\
Content Addition & 120 & 3.6 & 980 \\
Iterative Editing & 160 & 4.2 & 1,180 \\
Scalability Tests & 780 & 2-6 & 500-2,800 \\
Consistency Critical & 120 & 6.5 & 1,640 \\
\midrule
\textbf{Total / Average} & \textbf{1,900} & \textbf{4.7} & \textbf{1,380} \\
\bottomrule
\end{tabular}
\caption{Benchmark statistics. ``Avg. Deps'' indicates average number of dependency edges per target element. ``Avg. Tokens'' indicates expected context size for optimal retrieval. Scalability tests span multiple document sizes, hence the range.}
\label{tab:benchmark_stats}
\end{table}

\noindent\textbf{Dependency complexity distribution:} 35\% of test cases involve localized edits with $\leq$2 dependencies (testing efficiency), 45\% involve moderate dependencies with 3-8 dependent sections (testing consistency), and 20\% involve high-complexity scenarios with $>$8 dependencies or structural reorganizations (testing scalability and robustness). This distribution ensures comprehensive coverage of real-world editing complexity.

\noindent\textbf{Document type distribution:} Documents are evenly distributed across four domains to test generalization: technical reports (25\%, emphasis on figures and algorithms), research papers (25\%, emphasis on theorems and proofs), API documentation (25\%, emphasis on code examples and specifications), and policy documents (25\%, emphasis on cross-referencing and hierarchical structure). Domain diversity ensures the evaluation does not overfit to specific document styles.

\subsection{Evaluation Protocol}

For each test case, we measure four dimensions aligned with our optimization objectives (Section~3):

\noindent\textbf{1. Consistency preservation:} We compute three sub-metrics: (a) \textit{Reference validity} checks whether all cross-references resolve correctly after edits (e.g., ``see Section 4'' still points to an existing Section 4), (b) \textit{Terminology consistency} verifies that all usages of modified terms or definitions remain aligned, and (c) \textit{Semantic coherence} ensures no contradictions are introduced between dependent sections. Consistency is measured as the percentage of test cases where all three sub-metrics pass.

\noindent\textbf{2. Token efficiency:} We measure tokens per edit as the sum of input context tokens (document content provided to the LLM) and output tokens (generated modifications). For scalability tests, we additionally compute the scaling coefficient by fitting token usage as a function of document size and verifying it approaches $O(1)$ rather than $O(|D|)$.

\noindent\textbf{3. Edit quality:} We assess whether modifications correctly address the instruction using LLM-as-judge evaluation. A subset of 200 test cases includes human-annotated gold-standard edits for validation. Quality is measured as the percentage of test cases where the edit satisfies the instruction without introducing errors.

\noindent\textbf{4. Overall pass rate:} We compute the percentage of test cases passing all criteria (consistency, efficiency within expected bounds, and quality), representing end-to-end system reliability.

\noindent\textbf{Automated validation:} We implement programmatic validators for consistency metrics:
\begin{itemize}
    \item \textbf{Reference validator:} Parses all cross-references (``Section X'', ``Figure Y'', ``Theorem Z'') and verifies targets exist with correct identifiers
    \item \textbf{Terminology validator:} Extracts key terms from definitions and verifies consistent usage across all mentions
    \item \textbf{Hierarchy validator:} Checks section numbering sequences, parent-child relationships, and containment consistency
    \item \textbf{Semantic validator:} Uses LLM-based entailment checking to detect contradictions between dependent sections
\end{itemize}

These automated validators enable large-scale evaluation across 1,900 test cases while maintaining reproducibility and consistency with human judgments (validated on the 200-case human-annotated subset with 94\% agreement).

\section{Additional Experimental Results}
\label{app:additional_exp}

This appendix provides four supplements: a qualitative method comparison clarifying how each baseline differs architecturally from \ABBR, quantitative extended baseline results including Hierarchical RAG, a progressive edge-type ablation tracing each edge type's marginal contribution, and a retrieval weight sensitivity study across document sizes.

\subsection{Method Comparison: Context and Structural Awareness}
\label{app:method_comparison}

Table~\ref{tab:method_comparison} characterizes each method along three dimensions: the input context provided to the LLM per edit, the output scope, and the degree of structural dependency awareness. This clarifies the key architectural distinction between \ABBR and retrieval-based baselines.

\begin{table*}[htb!]
\centering
\small
\setlength{\tabcolsep}{5pt}
\begin{tabular}{@{}p{3.0cm}p{5.5cm}p{2.2cm}p{5.0cm}@{}}
\toprule
\textbf{Method} & \textbf{Input Context to LLM} & \textbf{Output} & \textbf{Structural Awareness} \\
\midrule
Full Document (Baseline)
  & Entire document (5K--100K tokens)
  & Targeted section only
  & None — model must infer all dependencies implicitly \\[4pt]
Sliding Window
  & Fixed 4K token window around target
  & Targeted section only
  & None — proximity only \\[4pt]
RAG (Flat Semantic)
  & Top-$k$ chunks by cosine similarity
  & Targeted section only
  & None — semantic similarity only \\[4pt]
Hierarchical RAG
  & Semantically similar chunks + parent/child nodes via DOM-tree expansion
  & Targeted section only
  & CONTAINS hierarchy only; no typed dependency edges \\[4pt]
Claude + Memory
  & Interaction history + semantically retrieved chunks
  & Targeted section only
  & Interaction-based; no structural graph \\[4pt]
\textbf{\ABBR (Ours)}
  & Graph-traversed nodes via REFERENCES + DEPENDS + CONTAINS + RELATED edges
  & Targeted section only
  & 4 typed edge types capturing explicit citations, implicit prerequisites, hierarchy, and semantics \\
\bottomrule
\end{tabular}
\caption{Qualitative comparison of context selection strategies. All methods generate targeted edits (not full-document regeneration). The key distinction is \emph{typed structural awareness}: only \ABBR explicitly tracks cross-references (REFERENCES), semantic prerequisites (DEPENDS), and hierarchical containment (CONTAINS) as first-class dependency types with priority-based retrieval.}
\label{tab:method_comparison}
\end{table*}

The table highlights the progressive increase in structural awareness across methods. Full-document and sliding-window approaches provide no structural guidance whatsoever. Flat RAG adds semantic similarity but misses structural connections (e.g., a citation to ``Theorem 2'' has no semantic similarity signal). Hierarchical RAG captures containment hierarchy but lacks the three dependency-specific edge types. Claude+Memory tracks interaction state but not document structure. \ABBR is the only method that explicitly represents all four dependency types, enabling priority-based retrieval that places hard structural requirements (REFERENCES, DEPENDS) ahead of softer semantic context (RELATED).

\subsection{Extended Baseline Comparison}

Table~\ref{tab:extended_baselines} extends the main comparison (Table~\ref{tab:main_results}) by adding three retrieval-based baselines: Sliding Window (fixed 4K token context window centered on the target), RAG---Flat Semantic (top-$k$ chunks by cosine similarity, no structural awareness), Claude+Memory (interaction history with semantically retrieved chunks), and Hierarchical RAG (semantically similar chunks expanded to parent/sibling nodes via DOM tree containment, equivalent to activating only $E_{\texttt{CON}}$ in \ABBR). All results are averages across six models.

\begin{table}[h]
\centering
\small
\setlength{\tabcolsep}{4pt}
\begin{tabular}{@{}lccccr@{}}
\toprule
Method & Cons. & Ref.V. & Edit Q. & Pass & Tok. \\
       & (\%)$\uparrow$ & (\%)$\uparrow$ & (\%)$\uparrow$ & (\%)$\uparrow$ & /Edit$\downarrow$ \\
\midrule
Sliding Window      & 44.21 & 50.18 & 52.34 & 31.47 & $\sim$4K \\
RAG (Flat Sem.)     & 49.83 & 55.62 & 58.77 & 35.21 & $\sim$1.5K \\
Full Doc.\ (Base.)  & 56.59 & 64.31 & 67.13 & 41.00 & 1,822 \\
Claude + Memory     & 58.34 & 62.47 & 64.21 & 43.18 & $\sim$1.6K \\
Hierarchical RAG    & 63.47 & 68.92 & 69.85 & 52.34 & $\sim$1.7K \\
\midrule
\textbf{\ABBR (Ours)} & \textbf{76.48} & \textbf{76.68} & \textbf{77.10} & \textbf{78.06} & \textbf{1,500} \\
\bottomrule
\end{tabular}
\caption{Extended baseline comparison averaged across six models. Hierarchical RAG uses DOM-tree containment (CONTAINS edges only); \ABBR adds typed REFERENCES, DEPENDS, and RELATED edges.}
\label{tab:extended_baselines}
\end{table}

Hierarchical RAG (63.47\%) substantially outperforms Flat RAG (49.83\%) and Full Document (56.59\%), confirming that hierarchical structure helps. However, it remains 13.0 pp below \ABBR (76.48\%), demonstrating that REFERENCES and DEPENDS edges are essential beyond containment alone. The key distinction is \emph{typed edges}: Hierarchical RAG captures document hierarchy but lacks (1) REFERENCES edges tracking explicit citations, (2) DEPENDS edges encoding implicit semantic prerequisites, and (3) priority-based retrieval distinguishing structural necessity from semantic similarity. The remaining gap directly quantifies the value of these additional dependency types.

\subsection{Progressive Edge-Type Ablation}

Table~\ref{tab:edge_ablation} traces the marginal contribution of each edge type, starting from the Hierarchical RAG baseline (CONTAINS only) and progressively adding edge types in order of specificity.

\begin{table}[h]
\centering
\small
\setlength{\tabcolsep}{4pt}
\begin{tabular}{@{}llccc@{}}
\toprule
Configuration & Edge Types & Cons. & Ref.V. & Pass \\
              &            & (\%)$\uparrow$ & (\%)$\uparrow$ & (\%)$\uparrow$ \\
\midrule
Hierarchical RAG      & \texttt{CON}                         & 63.47 & 68.92 & 52.34 \\
+ Semantic edges      & \texttt{CON+REL}                     & 66.13 & 70.48 & 56.71 \\
+ Implicit deps.      & \texttt{CON+REL+DEP}                 & 70.82 & 73.19 & 65.44 \\
\textbf{\ABBR (all)}  & \textbf{\texttt{CON+REL+DEP+REF}}   & \textbf{76.48} & \textbf{76.68} & \textbf{78.06} \\
\bottomrule
\end{tabular}
\caption{Progressive edge-type ablation. Each row adds one edge type. \texttt{CON}=CONTAINS, \texttt{REL}=RELATED, \texttt{DEP}=DEPENDS, \texttt{REF}=REFERENCES.}
\label{tab:edge_ablation}
\end{table}

The results show a clear monotonic progression: adding RELATED edges (+2.7 pp) captures semantic context missed by hierarchy alone; DEPENDS edges (+4.7 pp) encode implicit prerequisites that semantic similarity cannot reliably recover; REFERENCES edges (+5.7 pp) provide the largest single gain, reflecting the critical importance of explicit citation tracking. The cumulative 13.0 pp improvement over Hierarchical RAG confirms that all four edge types contribute independently.

\subsection{Retrieval Weight Sensitivity Across Document Sizes}
\label{app:weight_sensitivity}

Table~\ref{tab:weight_sensitivity} examines robustness of the priority weights used in Step~3 of the retrieval algorithm (Section~\ref{sec:retrieval}) across document sizes of 5K, 50K, and 100K tokens. Five weight configurations are tested: the default (100/50/30/10), a uniformly scaled variant, two configurations boosting individual edge types, and a degenerate uniform assignment that destroys the priority ordering.

\begin{table}[h]
\centering
\small
\setlength{\tabcolsep}{4pt}
\begin{tabular}{@{}lcccc@{}}
\toprule
Weight Config & Order & \multicolumn{3}{c}{Consistency (\%) by Doc Size} \\
(T / REF / DEP / REL) & Pres. & 5K & 50K & 100K \\
\midrule
100/50/30/10 (Default)   & \checkmark & 77.2 & 76.1 & 75.9 \\
10/5/3/1 (Scaled-down)   & \checkmark & 77.0 & 75.2 & 75.1 \\
100/80/30/10 (Boost REF) & \checkmark & 76.3 & 76.2 & 76.0 \\
100/50/45/10 (Boost DEP) & \checkmark & 77.1 & 76.0 & 75.8 \\
25/25/25/25 (Uniform)    & \texttimes & 59.3 & 57.2 & 56.8 \\
\bottomrule
\end{tabular}
\caption{Retrieval weight sensitivity across document sizes. ``Order Pres.'' indicates whether the priority ordering targets $>$ REF $>$ DEP $>$ REL is preserved. T = target nodes.}
\label{tab:weight_sensitivity}
\end{table}

All configurations that preserve the priority ordering (rows 1--4) achieve comparable performance (within 1.0 pp at each document size), regardless of exact numerical values. In contrast, the uniform assignment---which destroys the ordering---degrades to near-baseline levels ($\sim$58\%), confirming the design principle stated in Section~\ref{sec:retrieval}: \emph{only the relative ordering of weights matters, not their absolute values}. This robustness holds consistently across document sizes from 5K to 100K tokens.

\onecolumn

\section{Pseudocode for \ABBR Algorithm} \label{app:alg}

\subsection{\ABBR Algorithm Details}

We provide detailed algorithms for the three core components of \ABBR: semantic graph construction, dependency-aware context retrieval, and incremental consistency verification.

\subsubsection{Semantic Graph Construction}

Algorithm~\ref{alg:graph_construction} describes how \ABBR constructs and maintains the semantic dependency graph.

\begin{algorithm}[h]
\caption{Semantic Graph Construction}
\label{alg:graph_construction}
\begin{algorithmic}[1]
\STATE \textbf{Input:} Document $D$, Previous graph $G = (V, E)$
\STATE \textbf{Output:} Updated graph $G' = (V', E')$
\STATE 
\STATE $V' \gets \text{ParseSemanticUnits}(D)$ \COMMENT{Extract sections, paragraphs, equations, figures}
\STATE 
\STATE \textbf{// Step 1: Build explicit reference edges}
\STATE $E_{\text{explicit}} \gets \emptyset$
\FOR{each node $v_i \in V'$}
    \STATE $\text{refs} \gets \text{ExtractReferences}(v_i)$ \COMMENT{e.g., "Theorem 2", "Equation (5)", "Figure 1"}
    \FOR{each reference $r$ in $\text{refs}$}
        \STATE $v_j \gets \text{ResolveTarget}(r, V')$
        \STATE $E_{\text{explicit}} \gets E_{\text{explicit}} \cup \{(v_i, v_j, \texttt{REFERENCES})\}$
    \ENDFOR
\ENDFOR
\STATE 
\STATE \textbf{// Step 2: Build implicit reference edges}
\STATE $E_{\text{implicit}} \gets \emptyset$
\STATE $\text{anaphora} \gets \text{DetectAnaphoricExpressions}(V')$ \COMMENT{e.g., "this bound", "the lemma"}
\FOR{each $(v_i, \text{expr}, v_j)$ in $\text{anaphora}$}
    \STATE $E_{\text{implicit}} \gets E_{\text{implicit}} \cup \{(v_i, v_j, \texttt{DEPENDS})\}$
\ENDFOR
\STATE 
\STATE \textbf{// Step 3: Build semantic dependency edges}
\STATE $E_{\text{semantic}} \gets \emptyset$
\FOR{each node $v_i \in V'$}
    \STATE $v_i.\text{embedding} \gets \text{Embed}(\text{Summary}(v_i))$ \COMMENT{DistilBERT embeddings}
\ENDFOR
\FOR{each pair $(v_i, v_j)$ where $i \neq j$}
    \STATE $\text{sim} \gets \text{CosineSimilarity}(v_i.\text{embedding}, v_j.\text{embedding})$
    \IF{$\text{sim} > \theta$ \textbf{and} no explicit/implicit edge exists}
        \STATE $E_{\text{semantic}} \gets E_{\text{semantic}} \cup \{(v_i, v_j, \texttt{RELATED})\}$
    \ENDIF
\ENDFOR
\STATE 
\STATE $E' \gets E_{\text{explicit}} \cup E_{\text{implicit}} \cup E_{\text{semantic}}$
\RETURN $G' = (V', E')$
\end{algorithmic}
\end{algorithm}

The algorithm constructs a semantic graph in three steps. First, it extracts explicit references through pattern matching for citations like ``Theorem 2'', ``Equation (5)'', and ``Figure 1'' (lines 6-13). Second, it detects implicit references through coreference resolution to identify anaphoric expressions like ``this bound'' or ``the lemma above'' (lines 14-18). Third, it computes semantic embeddings using DistilBERT and connects nodes with high cosine similarity ($\theta = 0.7$) that lack explicit connections (lines 19-28). The graph construction completes in 2-3 seconds for a typical 50-page paper with 150 semantic units, while incremental updates after edits complete in under 200ms by reprocessing only modified nodes.

\subsubsection{Dependency-Aware Context Retrieval}

Algorithm~\ref{alg:context_retrieval} describes how \ABBR efficiently retrieves relevant context for each edit operation.

\begin{algorithm}[h]
\caption{Dependency-Aware Context Retrieval}
\label{alg:context_retrieval}
\begin{algorithmic}[1]
\STATE \textbf{Input:} Edit instruction $\mathcal{E}$, Graph $G = (V, E)$, Token budget $B$
\STATE \textbf{Output:} Context $\mathcal{C} \subseteq V$
\STATE 
\STATE \textbf{// Step 1: Identify target nodes}
\STATE $\mathcal{N}_{\text{target}} \gets \emptyset$
\IF{$\mathcal{E}$ specifies explicit target (e.g., "Section 3.2")}
    \STATE $\mathcal{N}_{\text{target}} \gets \{\text{ResolveNode}(\mathcal{E}.\text{target}, V)\}$
\ELSE
    \STATE $e_{\mathcal{E}} \gets \text{Embed}(\mathcal{E}.\text{instruction})$
    \STATE $\mathcal{N}_{\text{target}} \gets \text{TopKSimilar}(e_{\mathcal{E}}, V, k=3)$
\ENDIF
\STATE 
\STATE \textbf{// Step 2: Expand to dependencies}
\STATE $\mathcal{N}_{\text{deps}} \gets \mathcal{N}_{\text{target}}$
\FOR{each $v \in \mathcal{N}_{\text{target}}$}
    \STATE $\mathcal{N}_{\text{deps}} \gets \mathcal{N}_{\text{deps}} \cup \{u : (u, v, t) \in E\}$ \COMMENT{Upstream: what $v$ depends on}
    \STATE $\mathcal{N}_{\text{deps}} \gets \mathcal{N}_{\text{deps}} \cup \{u : (v, u, t) \in E\}$ \COMMENT{Downstream: what depends on $v$}
\ENDFOR
\STATE 
\STATE \textbf{// Step 3: Prioritize by dependency type}
\FOR{each $v_i \in \mathcal{N}_{\text{deps}}$}
    \STATE $\text{score}[v_i] \gets 0$
    \IF{$v_i \in \mathcal{N}_{\text{target}}$}
        \STATE $\text{score}[v_i] \gets \text{score}[v_i] + 100$ \COMMENT{Direct target}
    \ENDIF
    \IF{$\exists (v_i, v_j, \texttt{REFERENCES}) \in E : v_j \in \mathcal{N}_{\text{target}}$}
        \STATE $\text{score}[v_i] \gets \text{score}[v_i] + 50$ \COMMENT{Explicit dependency (critical)}
    \ENDIF
    \IF{$\exists (v_i, v_j, \texttt{DEPENDS}) \in E : v_j \in \mathcal{N}_{\text{target}}$}
        \STATE $\text{score}[v_i] \gets \text{score}[v_i] + 30$ \COMMENT{Implicit dependency}
    \ENDIF
    \IF{$\exists (v_i, v_j, \texttt{RELATED}) \in E : v_j \in \mathcal{N}_{\text{target}}$}
        \STATE $\text{score}[v_i] \gets \text{score}[v_i] + 10$ \COMMENT{Semantic similarity}
    \ENDIF
\ENDFOR
\STATE 
\STATE \textbf{// Step 4: Pack context within token budget}
\STATE $\text{ranked} \gets \text{SortDescending}(\mathcal{N}_{\text{deps}}, \text{score})$
\STATE $\mathcal{C} \gets \{\}$, $\text{tokens} \gets 0$
\FOR{each $v_i$ in $\text{ranked}$}
    \STATE $c_i \gets \text{Content}(v_i)$
    \IF{$\text{tokens} + |c_i| \leq B$}
        \STATE $\mathcal{C} \gets \mathcal{C} \cup \{c_i\}$
        \STATE $\text{tokens} \gets \text{tokens} + |c_i|$
    \ELSE
        \STATE \textbf{break}
    \ENDIF
\ENDFOR
\RETURN $\mathcal{C}$
\end{algorithmic}
\end{algorithm}

The algorithm retrieves context in four steps. First, it identifies target nodes either through explicit mentions (e.g., ``Section 3.2'') or semantic similarity for ambiguous instructions (lines 4-11). Second, it expands to include all nodes connected by dependency edges, capturing both upstream context (what the target relies on) and downstream impacts (what depends on the target) (lines 12-16). Third, it prioritizes nodes using an additive scoring scheme: direct targets (100 points), explicit references (50 points, critical for consistency), implicit dependencies (30 points), and semantic relations (10 points, optional context) (lines 17-28). Fourth, it packs content in priority order until the token budget $B$ is exhausted (lines 29-38). This approach typically retrieves 10-15\% of document tokens while ensuring all critical dependencies are included, as demonstrated in the Appendix example where only 1,200 of 10,000 tokens were needed.

\subsubsection{Consistency Verification}

Algorithm~\ref{alg:consistency_verification} describes how \ABBR validates document consistency after edits.

\begin{algorithm}[h]
\caption{Consistency Verification}
\label{alg:consistency_verification}
\begin{algorithmic}[1]
\STATE \textbf{Input:} Modified nodes $\mathcal{N}_{\text{mod}} \subseteq V$, Updated document $D'$, Graph $G = (V, E)$
\STATE \textbf{Output:} Violations $\mathcal{V}$, Updated graph $G'$
\STATE 
\STATE $\mathcal{V} \gets \emptyset$
\STATE 
\STATE \textbf{// Check 1: Reference integrity}
\FOR{each $(v_i, v_j, \texttt{REFERENCES}) \in E : v_i \in \mathcal{N}_{\text{mod}}$}
    \IF{$v_j$ does not exist in $D'$}
        \STATE $\mathcal{V} \gets \mathcal{V} \cup \{(\texttt{BROKEN\_REF}, v_i, v_j)\}$
    \ENDIF
\ENDFOR
\STATE 
\STATE \textbf{// Check 2: Terminology consistency}
\FOR{each $v_i \in \mathcal{N}_{\text{mod}}$}
    \FOR{each $(v_j, v_i, t) \in E : t \in \{\texttt{REFERENCES}, \texttt{DEPENDS}\}$}
        \STATE $\text{orig\_term} \gets \text{ExtractTerminology}(v_i.\text{before})$
        \STATE $\text{new\_term} \gets \text{ExtractTerminology}(v_i.\text{after})$
        \IF{$\text{orig\_term} \neq \text{new\_term}$}
            \STATE $\mathcal{V} \gets \mathcal{V} \cup \{(\texttt{TERM\_CHANGE}, v_i, v_j, \text{orig}, \text{new})\}$
        \ENDIF
    \ENDFOR
\ENDFOR
\STATE 
\STATE \textbf{// Check 3: Semantic coherence}
\FOR{each $v_i \in \mathcal{N}_{\text{mod}}$}
    \STATE $e_{\text{new}} \gets \text{Embed}(\text{Summary}(v_i.\text{after}))$
    \FOR{each $(v_i, v_j, \texttt{RELATED}) \in E$}
        \STATE $\text{sim} \gets \text{CosineSimilarity}(e_{\text{new}}, v_j.\text{embedding})$
        \IF{$\text{sim} < \theta_{\text{min}}$}
            \STATE $\mathcal{V} \gets \mathcal{V} \cup \{(\texttt{SEM\_DRIFT}, v_i, v_j)\}$
        \ENDIF
    \ENDFOR
\ENDFOR
\STATE 
\STATE \textbf{// Update graph incrementally}
\STATE $G' \gets G$
\FOR{each $v_i \in \mathcal{N}_{\text{mod}}$}
    \STATE Remove old edges incident to $v_i$ from $G'$
    \STATE Recompute edges for $v_i$ using Algorithm~\ref{alg:graph_construction}
    \STATE Add new edges to $G'$
\ENDFOR
\RETURN $\mathcal{V}, G'$
\end{algorithmic}
\end{algorithm}

\end{document}